\documentclass[11pt]{JHEP3}

\JHEPspecialurl{http://jhep.sissa.it/JOURNAL/JHEP3.tar.gz}

\usepackage{epsfig,multicol}
\usepackage{graphicx}
\usepackage{amsmath,amssymb}
\usepackage{mathrsfs}

\newcommand\fverb{\setbox\fverbbox=\hbox\bgroup\verb}
\newcommand\fverbdo{\egroup\medskip\noindent%
			\fbox{\unhbox\fverbbox}\ }
\newcommand\fverbit{\egroup\item[\fbox{\unhbox\fverbbox}]}
\newbox\fverbbox

\newcommand{\nn}{\nonumber}
\def\dfrac#1#2{\displaystyle\frac{#1}{#2}}

\newcommand{\pslash}{p\kern-1ex /}
\newcommand{\qslash}{q\kern-1ex /}
\newcommand{\lslash}{l\kern-1ex /}
\newcommand{\sslash}{s\kern-1ex /}
\newcommand{\kaslash}{k_a\kern-2ex /}
\newcommand{\kbslash}{k_b\kern-2ex /}
\newcommand{\Dslash}{\mathcal{D}\kern-1.5ex /}

\newcommand{\beqa}{\begin{eqnarray}}
\newcommand{\eeqa}{\end{eqnarray}}

\newcommand{\ba}{\begin{eqnarray}}
\newcommand{\ea}{\end{eqnarray}}
\newcommand{\be}{\begin{equation}}

\title{Extension of the HKLL bulk reconstruction for small $\Delta$}

\author{
$^a$Sinya Aoki\footnote{\tt saoki@yukawa.kyoto-u.ac.jp}\ \ and \ 
$^b$J\'anos Balog\footnote{\tt balog.janos@wigner.hu}
\\
\vskip 1ex
{\it $^a$Center for Gravitational Physics,\\
Yukawa Institute for Theoretical Physics, Kyoto University,\\
Kitashirakawa Oiwake-cho, Sakyo-Ku, Kyoto, Japan}  \\
\vskip 1ex
{\it $^b$Holographic QFT Group, Institute for Particle and Nuclear Physics,\\
Wigner Research Centre for Physics} \\
{\it H-1525 Budapest 114, P.O.B. 49, Hungary}\\
}

\received{\today} 		
\accepted{\today}	
\preprint{YITP-23-24}

\abstract{
We re-analyse the bulk reconstruction for a scalar field in Lorentzian AdS
spacetime, both for the case of even and odd dimensions, for an extended range
of conformal dimensions where the original HKLL reconstruction has to be
modified.
We also discuss the use of space-like Green's functions in the bulk
reconstruction. We demonstrate that in the extended range also the singular
part of the Green's function, omitted in the original papers, has be included.
The results are particularly simple and physically interesting for integer
conformal dimensions below the range considered in the original HKLL papers.
}

\begin{document}


\newcommand{\con}{\,\star\hspace{-3.7mm}\bigcirc\,}
\newcommand{\convu}{\,\star\hspace{-3.1mm}\bigcirc\,}
\newcommand{\Eps}{\Epsilon}
\newcommand{\gM}{\mathcal{M}}
\newcommand{\dD}{\mathcal{D}}
\newcommand{\gG}{\mathcal{G}}
\newcommand{\pa}{\partial}
\newcommand{\eps}{\epsilon}
\newcommand{\La}{\Lambda}
\newcommand{\De}{\Delta}
\newcommand{\nonu}{\nonumber}
\newcommand{\beq}{\begin{eqnarray}}
\newcommand{\eeq}{\end{eqnarray}}
\newcommand{\ka}{\kappa}
\newcommand{\ee}{\end{equation}}
\newcommand{\an}{\ensuremath{\alpha_0}}
\newcommand{\bn}{\ensuremath{\beta_0}}
\newcommand{\dn}{\ensuremath{\delta_0}}
\newcommand{\al}{\alpha}
\newcommand{\bm}{\begin{multline}}
\newcommand{\fm}{\end{multline}}
\newcommand{\de}{\delta}
\newcommand{\dpd}{\int {\rm d}^d p}
\newcommand{\dqd}{\int {\rm d}^d q}
\newcommand{\dxd}{\int {\rm d}^d x}
\newcommand{\dyd}{\int {\rm d}^d y}
\newcommand{\dud}{\int {\rm d}^d u}
\newcommand{\dzd}{\int {\rm d}^d z}
\newcommand{\dpp}{\int \frac{{\rm d}^d p}{p^2}}
\newcommand{\dqq}{\int \frac{{\rm d}^d q}{q^2}}


\section{Introduction}
The AdS/CFT correspondence \cite{Maldacena:1997re,Witten:1998qj}
relates a theory of gravity in AdS space to a conformal field
theory on the boundary.
One consequence of the correspondence is that bulk quantum fields
can be expressed as CFT operators.
In the large N limit the bulk fields are free and can be written
as smeared CFT operators.
The explicit construction, called the HKLL (Hamilton, Kabat,
Lifschytz, and Lowe)
bulk reconstruction,  was accomplished in a series of papers
\cite{Hamilton:2005ju,Hamilton:2006az,Kabat:2012hp}.
In the simplest case a massive free scalar field operator $\Phi(Y)$
is considered in ${\rm AdS}_{d+1}$.
The HKLL bulk reconstruction represents it in terms of the
boundary CFT primary of weight $\Delta$, $O(x)$, as
\beqa
\Phi(Y) &=& \int_{\Sigma_Y} {\rm d}x\, K(Y,x) O(x),
\label{eq:HKLL}
\eeqa
where $K(Y,x)$  is a smearing function, and the integration at the
boundary should be performed in a region  $\Sigma_Y$ space-like separated
from the bulk point $Y$. We refer to \cite{DeJonckheere:2017qkk,
Harlow:2018fse,Kajuri:2020vxf} for recent reviews.
See also \cite{Bhowmick:2019nso} for an alternative derivation based
on Gel'fand-Graev-Radon transforms.
Later the reconstruction has been extended to higher spins
as well\cite{Kabat:2012hp,Kabat:2012av,Heemskerk:2012np,
Kabat:2013wga,Sarkar:2014dma,Foit:2019nsr}. After having constructed the
free case the next step is to study bulk interactions \cite{Kabat:2011rz}.
An elegant way to introduce interactions as well as to reproduce the bulk
reconstruction for free fields is the method based on space-like Green's functions
\cite{Hamilton:2006az,Heemskerk:2012mn}.
  
In the original papers it was not explicitly stated that
\eqref{eq:HKLL} holds only for $\Delta > d-1$, due to the convergence for the integral.
This restriction is not essential for applications of the AdS/CFT correspondence in the case of
supersymmetric gauge theories, in particular in the prime example of the ${\cal N}=4$ SUSY U$(N)$
gauge theory in $d=4$ dimensions, 
since the conformal dimensions of
physically relevant operators are typically (much) larger
than this lower bound $\Delta > d-1$. See however \cite{Klebanov:1999tb} for some explicit examples for
small $\Delta$ primaries in the AdS/CFT context.
More importantly there exists an other family of models often
used in the AdS/CFT context, namely 
O$(N)$ vector models and their holographic
duals,  higher spin theories in the bulk\cite{Klebanov:2002ja,Sezgin:2002rt}.
In the most interesting $d=3$ case,  the simplest singlet operator
has $\Delta=1$ ($d-2$).
Furthermore, its square, an operator which can be used as a relevant deformation, has
$\Delta=2$ ($d-1$).
The HKLL formula \eqref{eq:HKLL} can not be used to relate
these singlet scalar operators in the free $O(N)$ vector model to bulk operators.

It turned out\cite{Kabat:2012hp} that for the special case $\Delta=d-1$
the smearing function in Poincare coordinates  is supported on the intersection of the light-cone of the bulk point and the
boundary. In \cite{DelGrosso:2019gow}
the range of allowed $\Delta$ values was extended down to $\Delta>d/2$  by analytic
continuation.
While the bulk-boundary relation remains linear,
the smearing kernel in (\ref{eq:HKLL}) is replaced by a suitable distribution.

In \cite{Aoki:2021ekk} we found, in some special cases mainly concentrating
on the (simpler) case of even AdS spaces, a
generalized HKLL formula for $\Delta$ values below the original lower
bound $d-1$ by a direct derivation, without using analytic continuation.
When we explicitly evaluated the results of
\cite {DelGrosso:2019gow}, we found that they precisely agree with the results
of the direct calculation in the range where they overlap. We also discussed the interesting special cases
$\Delta = d-s$, where $s$ is a positive integer only
limited by the requirement that the conformal weight satisfies the unitarity
bound $\Delta \ge( d-2)/2$ (equality  holds for the free scalar theory).
In these integer $\Delta$ cases the bulk operator $\Phi(Y)$ is expressed
in terms of CFT operators living on
$\Sigma_Y^{(0)}$ (boundary points light-like separated from $Y$).

In this paper we carefully re-analyse the HKLL bulk reconstruction, both for
the case of even and odd AdS spaces (odd and even boundary manifolds),
paying special attention to the range of conformal dimensions where the
construction is valid (not emphasized in the original HKLL papers).

After a setup for  the HKLL bulk reconstruction  in section~\ref{sec:setup}, we consider the case of even and odd ${\rm AdS}_{d+1}$
(odd and even $d$) in sections \ref{sec:odd_d} and \ref{sec:even_d}, respectively and first recall the very well-known HKLL bulk
reconstruction \cite{Hamilton:2005ju,Hamilton:2006az} for a massive
free scalar boson field with conformal weight $\Delta>d-1$ in each section.
The purpose of this
review is to introduce our notations and conventions,
which will be needed later in the paper when we extend the validity
of the construction to smaller values of $\Delta$. 
We then recall some pertinent results from \cite{Aoki:2021ekk} in both sections~\ref{sec:odd_d} and \ref{sec:even_d} before discussing
explicit reconstruction formulas for the regions $d-1>\Delta>d-2$ and
$d-2>\Delta>d-3$.  
Some detailed calculations and necessary properties are summarized in several appendices.
In addition, in appendices G and H, we discuss the use of space-like Green's functions in the bulk reconstruction.
We demonstrate that this alternative method correctly reproduces the same results but in the extended range also
the singular part of the Green's function, omitted in the original paper \cite{Hamilton:2006az}, has to be included.

\section{Setup for the HKLL bulk reconstruction in AdS$_{d+1}$} 
\label{sec:setup}
The HKLL bulk reconstruction\cite{Hamilton:2005ju,Hamilton:2006az} starts with a free scalar operator $\Phi(t,\rho, n)$ on the $d+1$ dimensional global  AdS spacetime, whose metric is given by
\beq
ds^2 &=&  R^2 d\rho^2 -R^2 \cosh^2\rho\, dt^2 +R^2\sinh^2\rho\, dn^i d n^i, 
\eeq
where $R$ is the AdS radius and $Y = (t,\rho, n^i)$ with $n\cdot n=1$ (or $Y=(t,\rho,\Omega)$)  are the standard global coordinates of AdS$_{d+1}$.

The value of $\Phi$ at the middle of the AdS, $Y_o = (0,0, n)$, is expressed as (see appendix~\ref{appA})
\beq
\Phi(Y_o) &=&  D(1) +   D_1(1), \quad D(z) = \sum_{n=0}^\infty d_n z^n, \  D_1(z) = \sum_{n=0}^\infty d_n^\dagger z^n,
\label{eq:bulk_op_middle}
\eeq
where $d_n$ and $d_n^\dagger$ are (rescaled) annihilation and creation operators, and $\Delta$ is related to the mass of the free scalar $m$ as
$m^2 R^2 =\Delta( \Delta-d)$.
On the other hand, using the BDHM relation 
\beq
O(x) &=& \lim_{\rho\to\infty} (\sinh\rho)^\Delta \Phi(\tilde t,\rho,\tilde n),  
\eeq
where $O(x)$ is a CFT operator with conformal dimension $\Delta$ at the AdS boundary $x=(\tilde t, \tilde n)$ with $\tilde n \cdot\tilde n=1$,
we have
\beqa
{\cal C}(\tilde t) &:=&\int{\rm d}\tilde\Omega\, O(\tilde t,\tilde \Omega) = e^{-i\Delta \tilde t} B(-e^{-2i\tilde t}) +  e^{i\Delta \tilde t} B_1(-e^{2i\tilde t}),
\label{eq:CFT_op} \\
&&B(z) := \sum_{n=0}^\infty b_n z^n, \quad  B_1(z) := \sum_{n=0}^\infty b_n^\dagger z^n,\
\eeq
and  $b_n, b_n^\dagger$ are related to $d_n,d_n^\dagger$ as
\beqa
b_n&:=& \Omega_d {P_n(1+\alpha)\over P_n(d/2)} d_n, \quad b_n^\dagger:= \Omega_d {P_n(1+\alpha)\over P_n(d/2)} d_n^\dagger,
\eeqa
where
$\alpha := \Delta-d/2$, $P_n(z):=\Gamma(z+n)/\Gamma(z)$ is the Pochhammer symbol, and $\Omega_d=\dfrac{2\pi^{d/2}}{\Gamma(d/2)}$
is the volume of the $d$ dimensional unit sphere.

The HKLL bulk reconstruction goes as follows. First a relation between $\Phi(Y_o)$ and $O(x)$ is derived, then $\Phi(Y)$ is obtained using the AdS
isometry $g$ and associated unitary operator $U(g)$ as $\Phi(Y) = U^\dagger (g) \Phi( Y_o) U(g)$, where $Y= g^{-1} Y_o$ is a generic point in the AdS space.

\section{Bulk reconstruction for odd $d$}
\label{sec:odd_d}
 We first consider the bulk reconstruction for the odd $d$ case.

\subsection{Results of the original HKLL bulk reconstruction}
In papers by HKLL\cite{Hamilton:2005ju,Hamilton:2006az}, a relation between $\Phi(Y_o)$ and $O(x)$ has been derived for odd $d$
(see also appendix~\ref{appX}) 
\beqa
\Phi(Y_o)&=& \xi \int {\cal D} x\,  k_0(\tilde t)
\Theta( \pi/2 -\tilde t) \Theta(\tilde t  +\pi/2) O(x),   
\label{eq:odd_HKLL}
\eeqa
where
\beqa
\xi := {1\over \pi\Omega_d}{\Gamma(1-d/2) \Gamma(1+\alpha)\over \Gamma(\nu+1)}, \quad
k_0(u):= (2 \cos u)^\nu, \quad \nu:=\Delta -d,  
\label{eq:xi}
\eeqa
The convergence of the $\tilde t$ integral near $\tilde t=\pm\pi/2$ implies $\nu > -1$. Thus the condition $\Delta > d-1$ is (implicitly) assumed for the original HKLL bulk reconstruction.

A relation at a generic point $Y=(t,\rho,n)$ was given as
\beq
\Phi (Y) &=& \xi \int {\cal D}x\, I^\nu(Y,x) T(Y,x) O(x), 
\label{eq:odd_HKLL_generic}
\eeq 
where
\beq
I(Y, x) &=& 2[\cosh(\rho) \cos(t-\tilde t) -(\sinh\rho) n^i \tilde n^i]\quad
T(Y,x) = \Theta(T_2-\tilde t) \Theta(\tilde t - T_1) ,
\eeq
where $ T_1=t-\omega, T_2= t+\omega$ with $\omega=\arccos[(\tanh\rho) n\cdot\tilde n]$ and $ 0<\omega<\pi$. Note that
\beq
\lim_{\rho,t\to 0} I^\nu(Y, x) &=&  k_0 (\tilde t),\quad \lim_{\rho,t\to 0} T(Y,x) = \Theta( \pi/2  -\tilde t) \Theta(\tilde t  +\pi/2).
\eeq

\subsection{Results of Ref. \cite{Aoki:2021ekk} for smaller $\Delta$}
\label{sec2-2}
In our previous paper\cite{Aoki:2021ekk} we have derived (see appendix~\ref{appY})
\beq
\Phi(Y_o) &=&{\eta\over 2\Omega_d} \left[ {\cal C}(\pi/2) + {\cal C}(-\pi/2) \right] + \xi \int_{\rm (sub)}{\rm d}\tilde t\,k_0(\tilde t) {\cal C}(\tilde t),
\label{eq:odd_ours}
\eeq
where 
\beq
\eta &=& {\Gamma(1-d/2)\Gamma(1+\alpha)\over \Gamma^2(1+\nu/2)}, 
\eeq
and the subtracted integral is defined by
\beqa
\int_{\rm (sub)}{\rm d}\tilde t\,K(\tilde t) f(\tilde t) &=& \int^0_{-\pi/2}{\rm d}\tilde t K(\tilde t)\left[f(\tilde t) -f(-\pi/2)\right]
+ \int^{\pi/2}_{0}{\rm d}\tilde t K(\tilde t)\left[f(\tilde t) -f(\pi/2)\right],~~~~~
\label{eq:int_subt}
\eeqa
which converges for $\nu > -2$ thanks to subtractions.
Thus \eqref{eq:odd_ours} is valid for $\Delta > d-2$ and it reduces to \eqref{eq:odd_HKLL} for $\Delta> d-1$.

For $\Delta = d-s$ with an integer $s$, simple relations without integral have been given \cite{Aoki:2021ekk}:
\beqa
\Phi(Y_o) &=&\xi_o{\prod_{k=1}^\ell \left\{{\partial^2\over \partial t^2} +(2k-1)^2\right\}\over \prod_{k=1}^{2\ell} (d-2k)} C_+(t)\big\vert_{t=0},\quad 
\xi_o := {(-1)^{{d-1\over 2}}\over 2\Omega_d}
\label{eq:ours_special}
\eeqa
for $\Delta=d-(2\ell+1)$, where $C_+(t) ={\cal C}(t+{\pi\over 2})+{\cal C}(t-{\pi\over 2})$, and
\beqa
\Phi(Y_o) &=&\xi_o{\prod_{k=1}^\ell \left\{{\partial^2\over \partial t^2} +4k^2\right\}\over \prod_{k=1}^{2\ell+1} (d-2k)}
\frac{\partial}{\partial t}C_-(t)\big\vert_{t=0}
\eeqa
for $\Delta=d-2(\ell+1)$, where $C_-(0) ={\cal C}(t+{\pi\over 2})-{\cal C}(t-{\pi\over 2})$.

In the previous paper we have not derived a formula for $\Phi(Y)$ at a generic point $Y$ for the whole extended range.
Results at the special points  $\Delta = d-1$ and $d-2$ only were given, which are as follows.
\beqa
\Phi(Y) &=&\xi_o \int{\rm d}\tilde\Omega\, {1\over {\cal R}(Y,x)}\left[O(T_1,\tilde\Omega) +O(T_2,\tilde\Omega)\right]
\label{eq:d-1}
\eeqa 
for $\Delta=d-1$, where ${\cal R}(Y,x) = \sqrt{\cosh^2\rho -(n\cdot \tilde n)^2\sinh^2\rho}$, and
\beqa
\Phi(Y) &=&\tilde \xi_o \int  {d\tilde\Omega\over {\cal R}^2(Y,x)}\left[\dot O(T_2,\tilde\Omega) -\dot O(T_1,\tilde\Omega)
  - \cot \omega\{O(T_2,\tilde\Omega) +O(T_1,\tilde\Omega)\}\right] ~~~
\label{eq:d-2}
\eeqa 
for $\Delta = d-2$,
where $\tilde\xi_o := {(-1)^{{d-1\over 2}}\over 2(d-2)\Omega_d}$, and $\dot O(x) :=\partial_{\tilde t} O(\tilde t, \tilde n)$.

\subsection{Bulk reconstruction for the extended range $\Delta > d-3$ for odd $d$}
In this subsection we derive a bulk reconstruction formula for a generic bulk point for $d-1 \geq \Delta > d-3$.

\subsubsection{Formula at a generic point by partial integration}
Since it is not easy to transform \eqref{eq:odd_ours} to a generic bulk point by the AdS isometry, we take a different strategy and we start from \eqref{eq:odd_HKLL_generic}, which is first rewritten by partial integration as
\beqa
\Phi(Y) = \eta_o (2\cosh\rho)^\nu \int{\rm d}\tilde\Omega&& \left\{ - {1\over  \Gamma(\nu+1)} \int_{-\omega}^\omega{\rm d}\tilde t\, \phi_1(\tilde t) \dot O(t+\tilde t,\tilde\Omega)\right. \nn \\ 
&&+\left.  K_1(\nu, \omega) \left[O(T_1,\tilde\Omega) + O(T_2,\tilde\Omega) \right] \right\},
\label{eq:generic_1}
\eeqa
where $\eta_o :=\Gamma(\nu+1)\xi = {\Gamma(1-d/2)\over \pi \Omega_d} \Gamma(1+\alpha)$, 
\beqa
\phi_1(u) &=&\int_0^u {\rm d}v\, \phi_0(v), \ \phi_0(u) := (\cos u -\cos \omega)^\nu,\
K_1(\nu,\omega) = {\phi_1(\omega)\over \Gamma(\nu+1)} .~~~~
\eeqa
Since $\phi_1(u) \sim (\omega-\vert u\vert)^{\nu+1}$ at $u\sim \pm \omega$, the $\tilde t$ integral is convergent for $\nu>-2$.

Performing a second integration by parts, \eqref{eq:generic_1} becomes
\beq
\Phi(Y) &=& \eta_o (2\cosh\rho)^\nu \int{\rm d}\tilde\Omega \left\{  {1\over  \Gamma(\nu+1)} \int_{-\omega}^\omega{\rm d}\tilde t\, \phi_2(\tilde t) \ddot O(t+\tilde t,\tilde\Omega)\right. \nn \\ 
&+&\left. 
K_2(\nu,\omega)\left[\dot O(T_1,\tilde \Omega) -\dot O(T_2,\tilde \Omega)\right]
+ K_1(\nu, \omega) \left[O(T_1,\tilde\Omega) + O(T_2,\tilde\Omega) \right] \right\},
\label{eq:generic_2}
\eeq
where
\beq
\phi_2(u) &=& \int_0^u\,{\rm d}v (u-v)\phi_0(v), \ \phi_2^\prime(u) = \phi_1(u), \
K_2(\nu,\omega)= {\phi_2(\omega)\over \Gamma(\nu+1)}.
\eeq
The $\tilde t$ integral in this expression is convergent for $\nu > -3$.

Although we do not need to go further for later analysis, 
we can repeat the procedure to obtain 
\beq
\Phi(Y) &=& \eta_o (2\cosh\rho)^\nu \int{\rm d}\tilde\Omega \left\{  {(-1)^k\over  \Gamma(\nu+1)} \int_{-\omega}^\omega{\rm d}\tilde t\, \phi_k(\tilde t)  O^{(k)}(t+\tilde t,\tilde\Omega)\right. \nn \\ 
&+&\left. 
\sum_{\ell=1}^k {\phi_\ell(\omega)\over \Gamma(\nu+1)}\left[ O^{(\ell-1)}(T_1,\tilde \Omega) +(-1)^{\ell-1} O^{(\ell-1)}(T_2,\tilde \Omega)\right]\right\}
\label{eq:generic_k}
\eeq 
for an arbitrary positive integer $k$, where
\beq
\phi_\ell(u) :={1\over (\ell-1)!}\int_0^u {\rm d}v \,(u-v)^{\ell-1}\phi_0(v), \quad O^{(\ell)}(u,\tilde \Omega) := {\partial^\ell\over \partial u^\ell} O(u,\tilde\Omega).
\eeq
 The $\tilde t$ integral is convergent for $ \nu > -(k+1)$.

\subsubsection{Analytic continuation of $K_1(\nu,\omega)$}
While the $\tilde t$ integral in \eqref{eq:generic_1}  is convergent for $ \nu > -2$,
we must show that $K_1(\nu,\omega)$ is convergent for $\nu > -2$. 
Using a limiting case of (3.663-1) in the table of integrals by Gradshteyn and Ryzhik, $K_1(\nu,\omega)$ can be evaluated  for $\nu > -1$ as
\beq
K_1(\nu, \omega) &=&\sqrt{\pi}2^\nu \left( \sin{\omega\over 2}\right)^{2\nu+1} {1\over \Gamma(\nu+3/2)}{}_2F_1\left({1\over 2},{1\over 2};\nu+{3\over 2};\sin^2{\omega\over 2}\right).
\label{eq:K1}
\eeq
Since the Gamma function in the denominator of \eqref{eq:K1} regularizes the hypergeometric function,   \eqref{eq:K1} 
can be analytically continued to all $\nu$.
The integral part is convergent for $\nu>-2$ and therefore \eqref{eq:generic_1} provides the analytic extension of the bulk reconstruction to the range $\nu > -2$.

\subsubsection{Analytic continuation of $K_2(\nu,\omega)$}
Since the $\tilde t$ integral in \eqref{eq:generic_2} is convergent for $\nu>-3$ and we have already seen that $K_1(\nu,\omega)$ is analytic for all $\nu$,
we now concentrate on the integral
\beq
K_2(\nu,\omega) =\omega K_1(\nu, \omega) -J_1(\nu,\omega),
\label{eq:K2}
\eeq
where
\beq
J_1(\nu,\omega) &:=& {1\over \Gamma(\nu+1)}\int_0^\omega {\rm d}u\, u (\cos u -\cos \omega)^\nu ,
\label{eq:Jint}
\eeq
which, unfortunately, can not be found in integral tables.

In the absence of an explicit formula for \eqref{eq:Jint} we have
derived (see appendix \ref{appD}) a
recursion relation for $J_1(\nu,\omega)$, which can also be used for
analytic continuation:
\begin{equation}
J_1(\nu,\omega)=\frac{1}{\sin^2\omega}\Big\{(\nu+2)^2\,
J_1(\nu+2,\omega)+(2\nu+3)\cos\omega \, J_1(\nu+1,\omega)+\frac
{(1-\cos\omega)^{\nu+2}}{\Gamma(\nu+3)}\Big\}.
\label{eq:recursion}
\end{equation}
The integrals related to the right hand side of (\ref{eq:recursion}) are convergent
for $\nu>-2$ and so this relation can be used to extend the left hand side
to $\nu>-2$ too. After this extension the right hand side will be defined
to $\nu>-3$ and it defines the left hand side also to $\nu>-3$. In this
way we can extend, step by step, $J_1(\nu,\omega)$ for all $\nu$.
Therefore, \eqref{eq:K2} implies that $K_2(\nu,\omega)$ is analytic for all $\nu$.
Since the $\tilde t$ integral is convergent for $\nu>-3$, and $K_{1,2}(\nu,\omega)$ are analytic for all $\nu$,
$\Phi(Y)$ in \eqref{eq:generic_2} provides the analytic extension of the bulk reconstruction to the range $\nu> -3$ ($\Delta > d-3$), as promised.

\subsubsection{Comparisons with previous results}
Since the starting formula in \eqref{eq:odd_HKLL_generic} is valid only for $\nu > -1$, 
results \eqref{eq:generic_1} for $\nu > -2$ and  \eqref{eq:generic_2} for $\nu > -3$ 
are not regarded as direct derivations but should be considered as analytic continuations to  $\nu > -2$ and $\nu > -3$.
Therefore, it is useful to compare \eqref{eq:generic_1} and \eqref{eq:generic_2} for special cases  with previous results obtained without using
analytic continuation.

For this purpose, using  the hypergeometric identities ${}_2F_1(a,b;c;z) =(1-z)^{-a} {}_2F_1(a,c-b;c;z/(z-1) )$,  
we rewrite \eqref{eq:K1} as
\beq
K_1(\nu, \omega) &=&\sqrt{\pi}2^\nu {\left( \sin{\omega\over 2}\right)^{2\nu+1}\over \cos{\omega\over 2}} {1\over \Gamma(\nu+3/2)}{}_2F_1\left({1\over 2},\nu+1;\nu+{3\over 2};-\tan^2{\omega\over 2}\right),
\label{eq:K1_alt}
\eeq
which gives 
\beq
K_1(-1,\omega) ={1\over \sin \omega} , \quad K_1(-2,\omega) = -{\cos \omega\over \sin^3 \omega}.
\eeq
To start the recursion of $J_1(\nu,\omega)$ for a negative integer $\nu$,
we calculate
\begin{equation}
J_1(0,\omega)=\frac{\omega^2}{2}, \quad  
J_1(1,\omega)=\int_0^\omega{\rm d}u\,u(\cos u-\cos \omega)=
-\frac{\omega^2}{2}\cos\omega+\cos\omega+\omega\sin\omega-1,
\end{equation}
which, through the recursion \eqref{eq:recursion}, lead to
\begin{equation}
J_1(-1,\omega)=\frac{\omega}{\sin\omega},\qquad  
J_1(-2,\omega)=\frac{1}{\sin^2\omega}-\frac{\omega\cos\omega}
{\sin^3\omega}.
\end{equation}
Thus, \eqref{eq:K2} leads to
\beqa
K_2(-1,\omega)=0, \quad K_2(-2,\omega)= -{1\over \sin^2 \omega} .
\eeqa

Using these,  \eqref{eq:generic_1} for $\nu=-1$ and \eqref{eq:generic_2} for $\nu=-2$ reduce to
\beq
\Phi(Y) &=&\xi_o \int {d\tilde\Omega\over {\cal R}(Y,x)} \left[ O(T_1,\tilde\Omega) + O(T_2,\tilde\Omega) \right],\\
\Phi(Y)&=&\tilde\xi_o \int \frac{{\rm d}\Omega}{{\cal R}^2(Y,x)}\Big\{
[{\dot O}(T_2,\Omega)-{\dot O}(T_1,\Omega)]-\cot\omega [O(T_2,\Omega)+O(T_1,\Omega)]\Big\},
\eeq
which reproduce our previous results from the direct evaluation, \eqref{eq:d-1} and \eqref{eq:d-2}.

By applying a (backward) partial integration  to the integral in  \eqref{eq:generic_1}, we obtain
\beq
\Phi(Y)&=&\eta_o(2\cosh\rho)^\nu\int{\rm d}\Omega\Big\{\frac{1}{\Gamma(\nu+1)}
\int_0^\omega{\rm d}u\,\phi(u)[O(t+u,\Omega)-O(T_2,\Omega)] + \frac{1}{\Gamma(\nu+1)}\nn \\
&\times&
\int_{-\omega}^0{\rm d}u\,\phi(u)[O(t+u,\Omega)-O(T_1,\Omega)]
+K_1(\nu,\omega)[O(T_2,\Omega)+O(T_1,\Omega)]\Big\}.
\eeq
For the middle point $Y_o$ this reduces to
\eqref{eq:odd_ours}, which was obtained by direct calculation for $\Delta > d -2$.

The above results show that the analytic continuation and the direct calculation without analytic continuation lead to the same formula
(at least in these special cases) for the extended range of $\Delta$.

\section{Bulk reconstruction for even $d$}
\label{sec:even_d}
We now consider a more difficult task, the extension of the bulk reconstruction for even $d$ to smaller $\Delta$.

\subsection{Results of the original HKLL bulk reconstruction}
For the even $d$ case, the result at the middle point has been obtained by HKLL\cite{Hamilton:2005ju,Hamilton:2006az}:
\beq
\Phi(Y_o) = \tilde\xi \int {\cal D}x\, T(Y_o,x)  I^\nu(Y_o,x) \ln[I(Y_o,x)] O(x), \quad  \tilde\xi:=\left(-{1\over \pi}\right)^{d/2+1} {\Gamma(1+\alpha)\over \Gamma(\nu+1)} .
\label{eq:HKLL_eve_mid}
\eeq

Using transformation properties under the AdS isometry $g$ 
\beqa
U^\dagger(g) \Phi(Y_o) U(g) = \Phi(g^{-1} Y_o), \quad U^\dagger(g) O(x) U(g) = H^\Delta(g^{-1},x) O(g^{-1}x), 
\eeqa
$\Phi$ at a generic bulk point $Y=g^{-1} Y_o$ becomes
\beqa
\Phi(Y) &=& \tilde\xi \int {\cal D}y\,T(Y_o,y) I^\nu(Y_o,y) \ln[I(Y_o,y)]  H^\Delta(g^{-1},y) O(g^{-1}y)\nn \\
&=& \tilde\xi \int {\cal D}x\,  T(g^{-1} Y_o,x) I^\nu(g^{-1} Y_o,x) \ln[I(g^{-1} Y_o,x) H(g,x) ]  O(x)\nn \\
&=& \Phi^{\rm HKLL}(Y) +\tilde\xi \hat\Phi(g),
\eeqa 
where
\beqa
 \Phi^{\rm HKLL}(Y) &=&\tilde\xi \int {\cal D}x\,  T(Y,x) I^\nu(Y,x) \ln[I(Y,x)  ]  O(x),
 \label{eq:even_HKLL}  \\
 \hat\Phi(g) &=& \int {\cal D}x\,  T(g^{-1}Y_o,x) I^\nu(g^{-1}Y_o,x) \ln[H(g,x) ]  O(x).
\eeqa
In the above derivation we used results in appendix \ref{appB} ((\ref{DD77}), (\ref{DD88}), (\ref{DD13}))
in the form
\beqa
I(Y_o,y) H(g^{-1},y) &=& I(g^{-1}Y_o, g^{-1} y), \quad
T(Y_o,y) = T(g^{-1}Y_o, g^{-1} y),
\eeqa
and
\beqa
{\cal D} (gx) H^d(g^{-1}, gx) = {\cal D} x, \quad H(g,x) ={1\over H(g^{-1}, gx)}.
\eeqa

It has been claimed in the HKLL papers\cite{Hamilton:2005ju,Hamilton:2006az}  that the bulk field at a generic point is given by $ \Phi^{\rm HKLL}(Y)$, which is true if
\beqa
\hat\Phi(g)\equiv 0
\label{Phig}  
\eeqa
for all group elements $g$.
Note that from the derivation it follows that $\hat\Phi(g)$ only depends on $Y=g^{-1}Y_o$. 
Although this was already discussed in appendix B of \cite{Hamilton:2006az},  
an elementary proof of (\ref{Phig}) is presented in appendix \ref{appC} for the sake of completeness. 

\subsection{Previous results  for integer $\Delta=d-s$}
In \cite{Aoki:2021ekk} we have derived results at the middle point for $\Delta = d -s$ with a positive integer $s$, which are summarized as
\beq
\Phi(Y_o) &=&\left. {(-1)^{d/2}\over \pi \Omega_d}{\prod_{k=1}^\ell \left\{ {\partial^2\over \partial t^2} +(2k-1)^2\right\}\over \prod_{k=1}^{2\ell} (d-2k)}
{\partial \over \partial\Delta} C_+(t)\right\vert_{t=0,\Delta=d-(2\ell+1)}, 
\label{eq:even1}\\
\Phi(Y_o) &=&\left. {(-1)^{d/2}\over \pi \Omega_d}{\prod_{k=1}^\ell \left\{ {\partial^2\over \partial t^2} +4k^2\right\}\over \prod_{k=1}^{2\ell+1} (d-2k)}
{\partial \over \partial t} {\partial \over \partial\Delta} C_-(t)\right\vert_{t=0,\Delta=d-2(\ell+1)}.
\label{eq:even2}
\eeq

\subsection{New results for smaller $\Delta$}
\subsubsection{Bulk reconstruction at the middle point}
\label{sec3-3-1}
For even $d$, the bulk reconstruction at the middle point $Y_o$ is given by
\begin{equation}
\Phi(Y_o)=\tilde\xi\int_{({\rm sub})}{\rm d}\tilde t\,k_1(\tilde t){\cal C}(\tilde t)
+\tilde\xi\,g^\prime(\nu)[{\cal C}(\pi/2)+{\cal C}(-\pi/2)].  
\label{finaleven}
\end{equation}
where
\beq
k_1(u) = (2\cos u)^\nu \ln (2\cos u), \quad g(\nu)=\int_0^{\pi/2} du\, (2\cos u)^\nu ={\pi\over 2}{\Gamma(\nu+1)\over \Gamma^2(\nu/2+1)}.
\eeq
Details of the derivation are presented in appendix \ref{appY}.

The derivative of the identity (\ref{extiden}) with respect to $\nu$ gives
\beq
\int_{({\rm sub})}{\rm d}\tilde t\,k_1(\tilde t){\cal C}(\tilde t) &+&  
\int_{({\rm sub})}{\rm d}\tilde t\,k_0(\tilde t){\cal C}_\Delta(\tilde t)\nn \\
&+&g^\prime(\nu)[{\cal C}(\pi/2)+{\cal C}(-\pi/2)]
+g(\nu)[{\cal C}_\Delta(\pi/2)+{\cal C}_\Delta(-\pi/2)]=0,~~
\eeq
which leads to an alternative form of the bulk reconstruction as
\begin{equation}
\Phi(Y_o)=-\tilde\xi\int_{({\rm sub})}{\rm d}\tilde t\,k_0(\tilde t){\cal C}_\Delta(\tilde t)
-\tilde\xi g(\nu)[{\cal C}_\Delta(\pi/2)+{\cal C}_\Delta(-\pi/2)],
\label{evenext}
\end{equation}
where we define
\beq
{\cal C}_\Delta(t) :={\partial \over \partial \Delta} {\cal C}(t)
\eeq

\subsubsection{Analytic continuation}
As in the odd $d$ case, analytic continuation is employed to obtain the bulk reconstruction at a generic point $Y$ also for even $d$.
We start with the formula $\Phi(Y) =\Phi^{\rm HKLL}(Y)$ in \eqref{eq:even_HKLL} rewritten as
\beq
\Phi(Y) &=& \tilde\eta (2\cosh\rho)^\nu \int{\rm d}\tilde\Omega\, H(t,\omega,\tilde\Omega), \quad \tilde \eta := \left(-{1\over \pi}\right)^{d/2+1} \Gamma(1+\alpha) ,\label{eq:Phi_from_H}
\eeq
where
\beq
H(t,\omega,\tilde\Omega) &:=& {1\over \Gamma(\nu+1)}\int_{-\omega}^\omega{\rm d}\tilde t\, g(\tilde t, \omega) O(\tilde t+t,\tilde\Omega), 
\eeq
with $g(u,\omega) :=(\cos u -\cos \omega)^\nu \ln (\cos u -\cos \omega)$.
A partial integration gives
\beqa
H(t,\omega,\tilde\Omega) &=& - {1\over \Gamma(\nu+1)}\int_{-\omega}^\omega{\rm d}\tilde t\, g_1(\tilde t, \omega) \dot O(\tilde t+t,\tilde\Omega)
+P_1(\nu,\omega) \left[ O(T_1,\tilde \Omega) +  O(T_2,\tilde \Omega)\right],~~~~~~~~
\label{eq:H}
\eeqa
where
\beq
g_1(u,\omega) &:=& \int_0^u dv\, g(v,\omega), \quad P_1(\nu,\omega): ={ g_1(\omega,\omega)\over \Gamma(\nu+1)},
\eeq

The $\tilde t$ integral in \eqref{eq:H} is convergent for $\nu > -2$. Although the integral defining $P_1(\nu,\omega)$ is convergent only for $\nu > -1$, 
it can be analytically continued using the recursion relation 
\beq
P_1(\nu,\omega) &=&{1\over \sin^2 \omega} \Bigl\{ (2\nu+3) \cos\omega P_1(\nu+1,\omega) +(\nu+2)^2 P_1(\nu+2,\omega) \nn \\
&+& 2\cos\omega K_1(\nu+1,\omega) +(\nu+2) K_1(\nu+2,\omega)\Bigr\} -{K_1(\nu,\omega)\over \nu+1},
\label{eq:recursionP}
\eeq
which is derived in appendix \ref{appD}.
The right hand side of the recursion relation is defined for $\nu> -2$ except a pole at $\nu=-1$, whose residue is given by
\beq
-K_1(-1,\omega) = -{1\over \sin \omega} .
\eeq
The recursion relation allows to extend $P_1(\nu,\omega)$ step by step to all $\nu$ but there will be poles for all negative integer values of $\nu$.

\subsubsection{Comparison at the middle point for $\Delta=d -1$}
While the presence of poles at negative integer $\nu$ prevents us from extending $\Phi(Y)$ to $\nu=-1$ by analytic continuation,
we can proceed for the special case $Y=Y_o$ (the middle point) starting from
\beq
\Phi(Y_o) &=& 2^\nu \tilde \eta \left\{  -{1\over \Gamma(\nu+1)}\int_{-{\pi\over 2}}^{\pi\over2}d\tilde t\, g_1\left(\tilde t,{\pi\over 2}\right) \dot {\cal C}(\tilde t) 
+ P_1\left(\nu,{\pi\over 2}\right)\left[ {\cal C}\left({\pi\over 2}\right) +  {\cal C}\left(-{\pi\over 2}\right) \right]
\right\}.~~~~~
\label{eq:Phi_middel}
\eeq
We then explicitly evaluate the integral (convergent for $\nu>-1$)
\beq
g_1\left({\pi\over 2},{\pi\over 2}\right)=\frac{\partial}{\partial\nu}[2^{-\nu}g(\nu)]=
 {\pi\over 2^{\nu+1}}{\Gamma(\nu+1)\over \Gamma^2(\nu/2+1)}\left\{ -\ln 2 +\psi(\nu+1) -\psi(\nu/ 2+1)\right\},
\eeq
giving
\beq
P_1\left(\nu,{\pi\over 2}\right) &=&  {\pi\over 2^{\nu+1}}{1\over \Gamma^2(\nu/2+1)}\left\{ -\ln 2 +\psi(\nu+1) -\psi(\nu/ 2+1)\right\}.
\label{eq:P1}
\eeq
Since the digamma function in \eqref{eq:P1} has a pole at $\nu = -1$  we see that
\beq
P_1 \left(\nu,{\pi\over 2}\right) = - {1\over \nu + 1} + O(1).
\eeq
On the other hand, since ${\cal C}(\tilde t)$ in \eqref{eq:CFT_op} is anti-periodic in $\tilde t$ with a period $\pi$ for an odd integer $\Delta$, 
we have
\beq
{\cal C}\left(t+{\pi\over 2}\right) +  {\cal C}\left(t-{\pi\over 2}\right) &=& (\nu +1) \left[{\cal C}_\Delta\left(t+{\pi\over 2}\right) +  {\cal C}_\Delta\left(t-{\pi\over 2}\right)\right] + O\left( (\nu+1)^2\right)
\eeq
near $\Delta = d+1$ for even $d$.  Thus we conclude that $\Phi(Y_o)$ is finite in the $\nu \to -1$ limit as
\beq
\Phi(Y_o) &=& {(-1)^{d/2}\over \pi \Omega_d}  \left[{\cal C}_\Delta\left({\pi\over 2}\right) +  {\cal C}_\Delta\left(-{\pi\over 2}\right)\right] ,
\label{eq:even_middle_int}
\eeq 
which reproduce the previous result in \eqref{eq:even1} for $\ell=0$, obtained by a different method without analytic continuation.

\subsubsection{An alternative expression at a generic point}
We can avoid difficulties arising from poles of $P_1(\nu, \omega)$ by considering an alternative expression of $\Phi$.

We start from the fact that for even $d$ and $\nu>-1$ (see (\ref{BB14})) 
\beq
\int {\cal D} x\, I^\nu(Y_0,x) T(Y_o,x) O(x) =0,
\eeq
which can be transformed to a generic point by the isometry transformation as
\beq
\int {\cal D} x\, I^\nu(Y,x) T(Y,x) O(x) =0 .
\eeq
By taking its derivative with respect to $\nu$ and comparing to (\ref{eq:even_HKLL}), we arrive at an alternative expression (valid for $\nu>-1$):
\beq
\Phi(Y)  &=& -\tilde\xi \int{\cal D} x\, I^\nu(Y,x) T(Y,x)  O_\Delta(x), \quad   O_\Delta(x):={\partial \over \partial \Delta} O(x).
\eeq
 
Applying the partial integration to this alternative expression, $H$ in  \eqref{eq:Phi_from_H} becomes 
\beqa
H(t, \omega,\tilde\Omega) &=& {1\over\Gamma(\nu+1)}\int_{-\omega}^\omega{\rm d}\tilde t\, \phi_1(\tilde t) \dot O_\Delta(\tilde t+t,\tilde\Omega) -K_1(\nu,\omega)\left[O_\Delta(T_1,\tilde\Omega) +O_\Delta(T_2,\tilde\Omega) \right].~~~~~~~~~
\label{eq:even_H}
\eeqa
 Since the $\tilde t$ integral is convergent for $\nu>-2$ and $K_1(\nu,\omega)$ is analytic for all $\nu$,  $\Phi(Y)$ can be analytically continued to 
 $\nu > -2$ as
 \beq
 \Phi(Y) &=&\tilde\eta (2\cos\rho)^\nu  \int{\rm d}\tilde\Omega\, H(t,\omega,\tilde \Omega)
 \label{eq:generic_even}
 \eeq
 with $H(t,\omega,\tilde \Omega)$ in \eqref{eq:even_H}.
 In particular for $\nu=-1$, we have
 \beq
 \Phi(Y)  &=& {(-1)^{d/2}\over \pi \Omega_d} \int{\rm d}\tilde \Omega {1\over {\cal R}(Y,x)} \left[O_\Delta(T_1,\tilde\Omega) +O_\Delta(T_2,\tilde\Omega) \right].
 \eeq
 For $Y_o$ this  agrees with \eqref{eq:even_middle_int}, and thus with \eqref{eq:even1} for $\ell=0$.
 
 For $Y_o$, but for generic $\nu>-2$ \eqref{eq:generic_even} reduces to
 \beq
 \Phi(Y_o) &=& -2^\nu\tilde\xi \int_{\rm (sub)}{\rm d}\tilde t\, (\cos (\tilde t))^\nu {\cal C}_\Delta(\tilde t)
 -2^\nu \tilde\eta K_1(\nu,\pi/2)\left[{\cal C}_\Delta(\pi/2) +{\cal C}_\Delta(-\pi/2) \right]\nn \\
 &=&- \tilde\xi \int_{\rm (sub)}{\rm d}\tilde t\, k_0(\tilde t)  {\cal C}_\Delta(\tilde t)
 -\tilde\xi g(\nu) \left[{\cal C}_\Delta(\pi/2) +{\cal C}_\Delta(-\pi/2) \right],
 \eeq
 reproducing \eqref{evenext}, which has been obtained without analytic continuation.

\vspace{5ex}
%

\section*{Acknowledgements}
We thank Dr. S. Terashima for useful discussions and for partially checking
our results.
J.B. acknowledges support from the International Research Unit of Quantum
Information (QIU) of Kyoto University Research Coordination Alliance,
and also would like to thank the Yukawa Institute for Theoretical Physics
at Kyoto University, where most of this work has been carried out, for
support and hospitality by the long term visitor program. 
This work has been supported in part by the NKFIH grant K134946, 
by the Grant-in-Aid of the Japanese Ministry of Education, Sciences and
Technology, Sports and Culture (MEXT) for Scientific Research
(Nos.~JP16H03978,  JP18H05236).

\par\bigskip

\appendix

\section{Fock space representation of the bulk and boundary fields}
\label{appA}
The construction of a massive bulk scalar field and the corresponding
boundary field with conformal weight $\Delta>d/2-1$ was reviewed in \cite{Aoki:2021ekk}.
Here we recall some formulas which will be used in this paper. 

A free bulk scalar field $\Phi$ in terms of  
canonical creation and annihilation operators
${\cal A}_{n\ell{\underline m}}^+$ and ${\cal A}_{n\ell{\underline m}}$
is represented as
\begin{equation}
\Phi(t,\rho,\Omega)=\sum_{n\ell{\underline m}}\sqrt{\frac{{\cal N}R}{2\nu_{n\ell}}}
\Big\{u_{n\ell}(\rho)Y_{\ell{\underline m}}(\Omega){\cal A}_{n\ell{\underline m}}
\,{\rm e}^{-i\nu_{n\ell} t}
+u_{n\ell}(\rho)Y_{\ell{\underline m}}(\Omega){\cal A}_{n\ell{\underline m}}^\dagger
\,{\rm e}^{i\nu_{n\ell} t}\Big\},    
\end{equation}
where ${\cal N}$ is a normalization constant related to the free Lagrangian,
$R$ is the AdS radius, $\nu_{n\ell}=\Delta+\ell+2n$ is the eigenfrequency,
$u_{n\ell}(\rho)$ is the radial wave function, and $Y_{\ell{\underline m}}(\Omega)$
are hyper-spherical harmonics for the $d-1$ dimensional sphere parametrized
alternatively by
the angular variables $\Omega$ or by the $d$ dimensional unit vector $n^i$. 
Note that for simplicity we are using real hyper-spherical harmonics.

The field at the middle point, \eqref{eq:bulk_op_middle}, can be expressed with the
rescaled Fock space operator,
\begin{equation}
d_n=\sqrt{\frac{{\cal N}R}{2\nu_{n0}}}\,(-1)^n \frac{P_n(d/2)}{n!}
{\cal N}_{n0}\frac{1}{\sqrt{\Omega_d}}{\cal A}_{n0{\underline 0}}.
\end{equation}
The explicit value of the normalization constant ${\cal N}_{n0}$ is not needed
in our calculation.

The BDHM relation \cite{Banks:1998dd} gives
the boundary field $O(\tilde t,\tilde \Omega)$ of conformal weight $\Delta$ as
\beqa
O(\tilde t,\tilde \Omega):=\lim_{\rho\to\infty}\left(\sinh\rho\right)^\Delta
\Phi(\tilde t,\rho,\tilde \Omega)
&=&\sum_{n\ell{\underline m}} \sqrt{\frac{{\cal N}R}{2\nu_{n\ell}}}
\Big\{{\rm e}^{-i\nu_{n\ell} \tilde t}\,\frac{P_n(1+\alpha)}{n!}
{\cal N}_{n\ell}Y_{\ell{\underline m}}(\tilde\Omega){\cal A}_{n\ell{\underline m}}\nn \\
&+&{\rm e}^{i\nu_{n\ell} \tilde t}\,\frac{P_n(1+\alpha)}{n!}{\cal N}_{n\ell}
Y_{\ell{\underline m}}(\tilde\Omega){\cal A}_{n\ell{\underline m}}^\dagger \Big\}.
\label{calO1}
\eeqa 

The integrated boundary field \eqref{eq:CFT_op} is given  
in terms of $b_n$, Fock space operators rescaled differently from $d_n$:
\begin{equation}
b_n=\sqrt{\frac{{\cal N}R}{2\nu_{n0}}}\,(-1)^n \frac{P_n(1+\alpha)}{n!}
{\cal N}_{n0}\sqrt{\Omega_d}{\cal A}_{n0{\underline 0}}
= \Omega_d\,\frac{P_n(1+\alpha)}{P_n(d/2)} d_n .
\label{ratio}
\end{equation}

\section{Reconstruction of the bulk field at the middle point}
\label{appX}

The relation \eqref{ratio}  leads to the basic formula for bulk reconstruction:
\beqa
D(w)&=&\frac{1}{2\pi i\Omega_d}\oint\frac{{\rm d}z}{z}B(z)\sum_{n=0}^\infty
\frac{P_n(d/2)}{P_n(1+\alpha)}\left(\frac{w} {z}\right)^n
=\frac{1}{2\pi i\Omega_d}\oint\frac{{\rm d}z}{z}B(z)
{}_2F_1(1,d/2;1+\alpha;w/z)\nn\\
&=&\frac{1}{2\pi i\Omega_d}\oint\frac{{\rm d}z}{z}B(w z)
{}_2F_1(1,d/2;1+\alpha;1/z),
\label{Dw1}
\eeqa
which is valid for an arbitrary physical $\Delta$ (except at $1+\alpha=0$).
Starting with \eqref{Dw1}, calculations go differently for odd and even $d$ (even and odd AdS). 

\subsection{Odd $d$}
\label{appX_odd}
For odd $d$ we can evaluate the integral \eqref{Dw1} using
the hypergeometric function identity (valid for odd $d$) 
\begin{equation}
{}_2F_1\left(1,{d\over 2};1+\alpha;{1\over z}\right)
=\frac{2\alpha z}{2-d}\,{}_2F_1\left(1,1-\alpha;2-{d\over 2};z\right)
+\frac{\Gamma(1-{d\over 2})\Gamma(1+\alpha)}{\Gamma(\nu+1)}
\left(-\frac{1}{z}\right)^{-{d\over 2}}(1-z)^{\nu},
\label{hyp}
\end{equation}  
where the first term is regular except for a cut starting at $z=1$. Around the
branch point $z=1$, its behavior is
\begin{equation}
{\rm regular}\ +\ {\rm const.}\,(1-z)^{\nu}.
\end{equation}  
When calculating the integral of this first term  in \eqref{Dw1}, we can
shrink our contour so
that it becomes a very small circle around the branch point $z=1$, and then,
its contribution vanishes for $\nu > -1$ ( {\it i.e.}  $\Delta > d-1$), because the value of the
integral gets smaller and smaller as our integral contour  gets smaller and
smaller.

The second term has a cut starting already at $z=0$. The contour can be shrunken
so that it becomes just the unit circle, since the singularity around the
second branch point $z=1$ is an integrable one for $\nu > -1$.
After a change of integration variable $z=-{\rm e}^{-2iu}$,
the integral along the unit circle becomes
\begin{equation}
D(w)=\frac{1}{\pi\Omega_d}\frac{\Gamma(1-d/2)\Gamma(1+\alpha)}
{\Gamma(\nu+1)}
\int_{-\pi/2}^{\pi/2}{\rm d}uB\left(-w{\rm e}^{-2iu}\right)
{\rm e}^{-i\Delta u}(2\cos u)^{\nu}.
\end{equation}  
Thus we obtain
\begin{equation}
D(1)=\xi
\int_{-\pi/2}^{\pi/2}{\rm d}u\,{\rm e}^{-i\Delta u} B\left(-{\rm e}^{-2iu}\right)
[2\cos(u)]^{\nu},
\end{equation}  
where $\xi$ is given by\eqref{eq:xi}.
If we repeat the whole calculation for $D_1$,  we have
\begin{equation}
D_1(1)=\xi
\int_{-\pi/2}^{\pi/2}{\rm d}u\,{\rm e}^{i\Delta u} B_1\left(-{\rm e}^{2iu}\right)
[2\cos(u)]^{\nu}.
\end{equation}
We can simply add the two contributions to arrive at \eqref{eq:odd_HKLL}.

\subsection{Even $d$}
For even $d$ our strategy is to make the calculation for generic, non-integer
dimension $d$ and take the limit $d\to{\rm even\ integer}$ at the end of the
calculation. The logic follows closely the HKLL paper \cite{Hamilton:2006az}.

We will use again the hypergeometric identity (\ref{hyp}),
which is valid for $d\not={\rm even\ integer}$. The first term on the
right hand side satisfies the identity
\begin{equation}
\begin{split}
{}_2F_1&\left(1,1-\alpha;2-\frac{d}{2};z\right)=\\
&\frac{1-\frac{d}{2}}{\nu}{}_2F_1(1,1-\alpha;1-\nu;1-z)
+\frac{\Gamma\left(2-\frac{d}{2}\right)\Gamma(-\nu)}{\Gamma(1-\alpha)}
z^{d/2-1}(1-z)^{\nu}.
\end{split}    
\label{hypid2}
\end{equation}
Here we have to require that $\nu\not={\rm integer}$, but this restriction
can be lifted soon. We will again  establish a relation between $D$ and $B$
(and between $D_1$ and $B_1$).

The starting point is again (\ref{Dw1}). Using the identity (\ref{hyp}) we 
notice again that the first term has no singularity inside the unit circle
and has a cut starting at 1 and therefore for this
term the integration contour can be shrunk to a small circle going around $z=1$.
Using the second hypergeometric identity (\ref{hypid2}) we see that the integral
along this small contour of radius $r$ is of the order O$(r^{\nu+1})$ and
therefore for $\nu> -1$ it vanishes in the limit $r\to0$.

Thus we are left with the second term in (\ref{hyp}) and the relation
(\ref{Dw1}) becomes
\begin{equation}
D(w)=\frac{\kappa}{2i}\Gamma\left(1-\frac{d}{2}\right)\oint\frac{{\rm d}z}{z}
B(w z)\left(-\frac{1}{z}\right)^{-d/2}(1-z)^{\nu},\quad  \kappa:=\frac{\Gamma(1+\alpha)}{\pi\Omega_d\Gamma(\nu+1)}.  
\label{DB}
\end{equation}
Similarly
\begin{equation}
D_1(w)=\frac{\kappa}{2i}\Gamma\left(1-\frac{d}{2}\right)\oint\frac{{\rm d}z}{z}
B_1(w z)\left(-\frac{1}{z}\right)^{-d/2}(1-z)^{\nu}.
\label{D1B1}
\end{equation}
The above formulas are apparently divergent for even $d$ because of the singular
factor $\Gamma(1-d/2)$ but we can observe that the integrals
\begin{equation}
\oint\frac{{\rm d}z}{z}
B(w z)\left(-\frac{1}{z}\right)^{-d_o/2}(1-z)^{\nu}, \quad \oint\frac{{\rm d}z}{z} B_1(w z)\left(-\frac{1}{z}\right)^{-d_o/2}(1-z)^{\nu}
\label{vanB}
\end{equation}
vanish for even $d_o$. The reason is the same as above: in this case the
integrand has no singularity (or cut starting) at $z=0$ and therefore the
integration contour can again be shrunk to a small circle around $z=1$ and the
integral vanishes for $\nu > -1$.

Using this observation we can proceed as follows. We write $d=d_o+\varepsilon$
and subtract the vanishing integral from (\ref{DB}). Reducing the contour to the
unit circle and using the integration variable $z=-e^{-2iu}$, (\ref{DB}) can be
written as
\begin{equation}
D(w)=\kappa\Gamma(1-d/2)\int_{-\pi/2}^{\pi/2}{\rm d}uB\left(-we^{-2iu}\right)
e^{-iu\Delta}(2\cos u)^{\nu}\left(1-e^{iu\varepsilon}\right)  
\end{equation}
and a similar expression for $D_1(w)$.
Now the O$(1/\varepsilon)$ prefactors are compensated by the fact that the
integrals are O$(\varepsilon)$ and in the limit $d\to d_o$ we get
\begin{equation}
D(w)=\frac{2\kappa(-1)^{d_o/2}}{\Gamma(d_o/2)}\int_{-\pi/2}^{\pi/2}{\rm d}u
B\left(-we^{-2iu}\right)
e^{-iu\Delta}(2\cos u)^{\nu_o}(-iu) , \quad \nu_o :=\Delta-d_o 
\label{minusiu}
\end{equation}
and
\begin{equation}
D_1(w)=\frac{2\kappa(-1)^{d_o/2}}{\Gamma(d_o/2)}\int_{-\pi/2}^{\pi/2}{\rm d}u
B\left(-we^{2iu}\right)
e^{iu\Delta}(2\cos u)^{\nu_o}(iu).  
\label{plusiu}
\end{equation}
Using integration along the unit circle, the vanishing first integral in (\ref{vanB})
can be written as the identity
\begin{equation}
0=\int_{-\pi/2}^{\pi/2}{\rm d}uB\left(-we^{-2iu}\right)
e^{-iu\Delta}(2\cos u)^{\nu_o}.  
\label{BB14}
\end{equation}
Since this is true for all $\Delta$ for $\nu_o > -1$, following HKLL
\cite{Hamilton:2006az}, we take its derivative with respect to $\Delta$, which
gives
\begin{equation}
0=\int_{-\pi/2}^{\pi/2}{\rm d}uB\left(-we^{-2iu}\right)
e^{-iu\Delta}(2\cos u)^{\nu_o}[\ln(2\cos u)-iu].  
\end{equation}
Using this identity and putting $w=1$ we obtain from (\ref{minusiu})
\begin{equation}
D(1)=-\frac{2\kappa(-1)^{d_o/2}}{\Gamma(d_o/2)}\int_{-\pi/2}^{\pi/2}{\rm d}u
B\left(-e^{-2iu}\right)
e^{-iu\Delta}(2\cos u)^{\nu_o}\ln(2\cos u),
\label{Dat1}
\end{equation}
and similarly
\begin{equation}
D_1(1)=-\frac{2\kappa(-1)^{d_o/2}}{\Gamma(d_o/2)}\int_{-\pi/2}^{\pi/2}{\rm d}u
B_1\left(-e^{2iu}\right)
e^{iu\Delta}(2\cos u)^{\nu_o}\ln(2\cos u).  
\label{D1at1}
\end{equation}
Now the two terms can be simply added and we get the final result \eqref{eq:HKLL_eve_mid},
where we dropped the subscript $o$ from the dimension $d$,
which is from now on an even integer again.

\section{Bulk reconstruction for $\Delta>d-2$ (middle point)}
\label{appY}
In this appendix we give the details of the derivation of the bulk
reconstruction formulas for the middle point, both for the odd and even $d$
cases, in the extended range $\Delta>d-2$. 
To begin with, we rewrite (\ref{Dw1}) by adding
and subtracting $B(w)$ under the integral as
\beqa
D(w)
&=&\frac{B(w)}{2\pi i\Omega_d}\oint\frac{{\rm d}z}{z}\,
{}_2F_1\left(1,{d\over 2};1+\alpha;{1\over z}\right)\nn\\
&+&\frac{1}{2\pi i\Omega_d}\oint\frac{{\rm d}z}{z}[B(w z)-B(w)]\,
{}_2F_1\left(1,{d\over 2};1+\alpha;{1\over z}z\right) .
\label{32}
\eeqa

\subsection{Odd $d$}
For odd $d$, using (\ref{32}), the manipulations in section~\ref{appX_odd} 
remain valid for the extended range
$\Delta>d-2$ ($\nu > -2$) and we obtain
\begin{equation}
D(w)=\frac{B(w)}{\Omega_d}+\xi
\int_{-\pi/2}^{\pi/2}{\rm d}u\, {\rm e}^{-iu\Delta}[2\cos(u)]^{\nu}
\{B(-w{\rm e}^{-2iu})-B(w)\},
\label{Dw}
\end{equation}
where singularities near $u=\pm {\pi\over 2}$ of the integrand become
integrable for $\nu > -2$ thanks to the subtraction of $B(w)$.

We then separate the integrated boundary field
${\cal C}(t)$ into positive/negative frequency parts
${\cal C}_+(t)/{\cal C}_-(t)$, which are given by the two terms of \eqref{eq:CFT_op}. 
Using these definitions, we have the identity
\begin{equation}
{\rm e}^{-i\Delta t}B({\rm e}^{-2it})={\rm e}^{-i\frac{\Delta\pi}{2}}
{\cal C}_+(t-\pi/2)={\rm e}^{i\frac{\Delta\pi}{2}}{\cal C}_+(t+\pi/2).
\end{equation}
Thus (\ref{Dw}) leads to
\beqa
D(1)&=&\frac{1}{\Omega_d}{\rm e}^{-\frac{i\Delta\pi}{2}}{\cal C}_+(-\pi/2)
+\xi\int_{-\pi/2}^0{\rm d}u [2\cos(u)]^{\nu}\{{\cal C}_+(u)
-{\rm e}^{-i(u+\pi/2)\Delta}{\cal C}_+(-\pi/2)\}\nn \\
&+&\xi\int_0^{\pi/2}{\rm d}u [2\cos(u)]^{\nu}\{{\cal C}_+(u)
-{\rm e}^{-i(u-\pi/2)\Delta}{\cal C}_+(\pi/2)\}.
\eeqa
Next by adding and subtracting an integral proportional to $k_+$ for the first
integral and $k_-$ for the second integral, where
\begin{equation}
k_\pm=\xi\int_0^{\pi/2}{\rm d}u(2\cos u)^{\nu}
[1-{\rm e}^{\pm i\Delta(u-\pi/2)}],
\end{equation}
which are convergent for $\nu>-2$, we obtain
\beq
&&D(1)=\left\{\frac{1}{\Omega_d}+{\rm e}^{\frac{i\Delta\pi}{2}} k_+
+{\rm e}^{\frac{-i\Delta\pi}{2}} k_-\right\}
{\rm e}^{-\frac{i\Delta\pi}{2}}{\cal C}_+(-\pi/2)\nn \\
&+&\xi\int_{-\pi/2}^0{\rm d}u [2\cos u]^{\nu}\{{\cal C}_+(u)
-{\cal C}_+(-\pi/2)\}+\xi\int_0^{\pi/2}{\rm d}u [2\cos u]^{\nu}\{{\cal C}_+(u)
-{\cal C}_+(\pi/2)\}.~~~~~~
\eeq
Analogously, repeating the calculation with $D_1$ and ${\cal C}_-$, we have
\begin{equation}
{\rm e}^{i\Delta t}B_1({\rm e}^{2it})={\rm e}^{i\frac{\Delta\pi}{2}}
{\cal C}_-(t-\pi/2)={\rm e}^{-i\frac{\Delta\pi}{2}}{\cal C}_-(t+\pi/2)
\end{equation}
and
\beqa
&&D_1(1)=\left\{\frac{1}{\Omega_d}+{\rm e}^{\frac{i\Delta\pi}{2}} k_+
+{\rm e}^{\frac{-i\Delta\pi}{2}} k_-\right\}
{\rm e}^{\frac{i\Delta\pi}{2}}{\cal C}_-(-\pi/2)\nn \\
&+&\xi\int_{-\pi/2}^0{\rm d}u [2\cos u]^{\nu}\{{\cal C}_-(u)
-{\cal C}_-(-\pi/2)\}
+\xi\int_0^{\pi/2}{\rm d}u [2\cos u]^{\nu}\{{\cal C}_-(u)
-{\cal C}_-(\pi/2)\}.~~~~~~~
\eeq

These results can be further simplified by using the following two identities.
\begin{equation}
\begin{split}
{\cal C}(t&-\pi/2)+{\cal C}(t+\pi/2)
={\cal C}_+(t-\pi/2)+{\cal C}_-(t-\pi/2)+{\cal C}_+(t+\pi/2)+
{\cal C}_-(t+\pi/2)\\
&=2\cos\frac{\Delta\pi}{2}\big\{{\rm e}^{-\frac{i\Delta\pi}{2}}{\cal C}_+(t-\pi/2)+
{\rm e}^{\frac{i\Delta\pi}{2}}{\cal C}_-(t-\pi/2)\big\},
\end{split}
\end{equation}
\begin{equation}
\begin{split}
\int_0^{\pi/2}{\rm d}u&(2\cos u)^A\left\{\cos\frac{B\pi}{2}-\cos Bu\right\}\\
&=\frac{\pi}{2}\,\Gamma(1+A)\Big\{\frac{\cos\frac{B\pi}{2}}{\Gamma^2(1+A/2)}
-\frac{1}{\Gamma(1+\frac{A-B}{2})\Gamma(1+\frac{A+B}{2})}\Big\}
\end{split}
\end{equation}
for $A>-2$, $B=A+d$, $d=3,5,7,\cdots$.
Using the second identity we find
\begin{equation}
\begin{split}
{\rm e}^{\frac{i\Delta\pi}{2}}k_+&+{\rm e}^{-\frac{i\Delta\pi}{2}}k_-=
\xi\int_0^{\pi/2}{\rm d}u(2\cos u)^{\Delta-d}\big\{2\cos\frac{\Delta\pi}{2}-
2\cos\Delta u\big\}
=\frac{1}{\Omega_d}\big\{\eta\cos\frac{\Delta\pi}{2}-1\big\}.
\end{split}
\end{equation}
Finally, adding $D(1)$ and $D_1(1)$ we obtain the final result \eqref{eq:odd_ours}.

\subsection{Even $d$}
For the even $d$ case by adding and subtracting a constant we rewrite the
starting formula (\ref{Dw1}) as
\begin{equation}
D(1)=\frac{B(1)}{\Omega_d}+\frac{1}{2\pi i\Omega_d}\oint\frac{{\rm d}z}{z}
B_o(z)(1-z)\,{}_2F_1(1,d/2;1+\alpha;1/z)  
\end{equation}
and similarly
\begin{equation}
D_1(1)=\frac{B_1(1)}{\Omega_d}+\frac{1}{2\pi i\Omega_d}\oint\frac{{\rm d}z}{z}
B_{1o}(z)(1-z)\,{}_2F_1(1,d/2;1+\alpha;1/z),  
\end{equation}
where we introduced the formally holomorphic fields $B_o(z)$, $B_{1o}(z)$ by
\begin{equation}
\hat B(z)=B(z)-B(1)=(1-z)B_o(z),\quad
\hat B_1(z)=B_1(z)-B_1(1)=(1-z)B_{1o}(z).
\end{equation}
We also introduce $\Delta_o=\Delta+1$ and note that $\Delta_o>d-1$.

Now we can copy the results of our calculation valid for the original range
($\Delta>d-1$) with the following modifications: there is the extra constant
term for both $D(1)$ and $D_1(1)$, the role of $B(z)$ and $B_1(z)$ is played
by $B_o(z)$ and $B_{1o}(z)$, respectively, and we put $\Delta_o$ in place of
$\Delta$. We obtain
\begin{equation}
D(1)=\frac{B(1)}{\Omega_d}+\frac{\kappa}{2i}\Gamma(1-d/2)
\oint\frac{{\rm d}z}{z}
B_o(z)\left(-\frac{1}{z}\right)^{-d/2}(1-z)^{\Delta_o-d}  
\end{equation}
and
\begin{equation}
D_1(1)=\frac{B_1(1)}{\Omega_d}+\frac{\kappa}{2i}\Gamma(1-d/2)
\oint\frac{{\rm d}z}{z}
B_{1o}(z)\left(-\frac{1}{z}\right)^{-d/2}(1-z)^{\Delta_o-d}.  
\end{equation}
At this point we still have to regularize the dimension as $d=d_o+\epsilon$,
where $d_o$ is an even integer. After carrying out the $\epsilon\to0$ limit and
restoring $\hat B(z)$ and $\hat B_1(z)$ we find
\begin{equation}
D(1)=\frac{B(1)}{\Omega_d}-\frac{2\kappa(-1)^{d_o/2}}{\Gamma(d_o/2)}
\int_{-\pi/2}^{\pi/2}{\rm d}u\,\hat B\left(-e^{-2iu}\right)e^{-iu\Delta}
(2\cos u)^{\nu_o}\ln(2\cos u)
\end{equation}
and
\begin{equation}
D_1(1)=\frac{B_1(1)}{\Omega_d}-\frac{2\kappa(-1)^{d_o/2}}{\Gamma(d_o/2)}
\int_{-\pi/2}^{\pi/2}{\rm d}u\,\hat B_1\left(-e^{2iu}\right)e^{iu\Delta}
(2\cos u)^{\nu_o}\ln(2\cos u).
\end{equation}
There are also identities of the form
\begin{equation}
\int_{-\pi/2}^{\pi/2}{\rm d}u\,\hat B\left(-e^{-2iu}\right)e^{-iu\Delta}
(2\cos u)^{\nu_o}=
\int_{-\pi/2}^{\pi/2}{\rm d}u\,\hat B_1\left(-e^{2iu}\right)e^{iu\Delta}
(2\cos u)^{\Delta-d_o}=0 .
\label{evenid1}
\end{equation}
From now on we drop the subscript from $d_o$ and use the notation $d$ for the
dimension.

The identities (\ref{evenid1}) take the form
\begin{equation}
\int_{-\pi/2}^{\pi/2}{\rm d}u\,k_0(u)\left\{{\cal C}_+(u)-B(1)e^{-iu\Delta}\right\}
= \int_{-\pi/2}^{\pi/2}{\rm d}u\,k_0(u)\left\{{\cal C}_-(u)-B_1(1)e^{iu\Delta}\right\}
=0,  
\end{equation}
whereas $D(1)$, $D_1(1)$ are given by
\begin{equation}
D(1)=\frac{B(1)}{\Omega_d}+\tilde\xi
\int_{-\pi/2}^{\pi/2}{\rm d}u\,k_1(u)\left\{{\cal C}_+(u)-B(1)e^{-iu\Delta}\right\},
\end{equation}
\begin{equation}
D_1(1)=\frac{B_1(1)}{\Omega_d}+\tilde\xi
\int_{-\pi/2}^{\pi/2}{\rm d}u\,k_1(u)\left\{{\cal C}_-(u)
-B_1(1)e^{iu\Delta}\right\}.
\end{equation}
We recall that
\begin{equation}
\tilde\xi=-\frac{2\kappa(-1)^{d/2}}{\Gamma(d/2)}=
\left(-\frac{1}{\pi}\right)^{1+d/2}\,\frac{\Gamma(1+\alpha)}
{\Gamma(\nu+1)}.
\end{equation}

By adding and subtracting terms we can show that
\begin{equation}
\begin{split}  
\int_{-\pi/2}^{\pi/2}&{\rm d}u\,k_i(u)\left\{{\cal C}_+(u)-B(1)e^{-iu\Delta}
\right\}\\
&=\int_{({\rm sub})}{\rm d}u\,k_i(u){\cal C}_+(u)+2B(1)\int_0^{\pi/2}
{\rm d}u\,k_i(u)\left(\cos\frac{\pi\Delta}{2}-\cos u\Delta\right)
\end{split}  
\end{equation}
and similarly
\begin{equation}
\begin{split}  
\int_{-\pi/2}^{\pi/2}&{\rm d}u\,k_i(u)\left\{{\cal C}_-(u)-B_1(1)e^{iu\Delta}
\right\}\\
&=\int_{({\rm sub})}{\rm d}u\,k_i(u){\cal C}_-(u)+2B_1(1)\int_0^{\pi/2}
{\rm d}u\,k_i(u)\left(\cos\frac{\pi\Delta}{2}-\cos u\Delta\right).
\end{split}  
\end{equation}
Adding the two identities written in this form we get
\begin{equation}
\int_{({\rm sub})}{\rm d}u\,k_0(u){\cal C}(u)+2\left[B(1)+B_1(1)\right]
\int_0^{\pi/2}
{\rm d}u\,k_0(u)\left(\cos\frac{\pi\Delta}{2}-\cos u\Delta\right)=0.
\end{equation}
The bulk field at the middle point is given by
\begin{equation}
\begin{split}
\Phi(Y_o)&=\tilde\xi\int_{({\rm sub})}{\rm d}u\,k_1(u){\cal C}(u)\\
&+\left[B(1)+B_1(1)\right]\Big\{\frac{1}{\Omega_d}+
2\tilde\xi\int_0^{\pi/2}
{\rm d}u\,k_1(u)\left(\cos\frac{\pi\Delta}{2}-\cos u\Delta\right)\Big\}.
\end{split}
\label{fin0}
\end{equation}
To calculate the remaining integrals we will use 
\begin{equation}
\int_0^{\pi/2}{\rm d}u\,(2\cos u)^A\left(\cos\frac{\Delta\pi}{2}-\cos u\Delta
\right)=\cos\frac{\Delta\pi}{2}\,g(A)-\frac{\pi}{2}
\frac{\Gamma(1+A)}{\Gamma\left(1+\frac{A-\Delta}{2}\right)
\Gamma\left(1+\frac{A+\Delta}{2}\right)},
\end{equation}
which leads to
\begin{equation}
\int_0^{\pi/2}{\rm d}u\,k_0(u)\left(\cos\frac{\Delta\pi}{2}-\cos u\Delta\right)
=\cos\frac{\Delta\pi}{2}\,g(\nu),  
\end{equation}
since the second term does not contribute in this case. Thus the final form
of the identity becomes
\begin{equation}
\int_{({\rm sub})}{\rm d}u\,k_0(u){\cal C}(u)+g(\nu)\,
\big({\cal C}(\pi/2)+{\cal C}(-\pi/2)\big)=0.
\label{extiden}
\end{equation}

Next we calculate
\begin{equation}
\begin{split}  
\int_0^{\pi/2}{\rm d}u\,k_1(u)&\left(\cos\frac{\Delta\pi}{2}-\cos u\Delta
\right)=\cos\frac{\Delta\pi}{2}\,g^\prime(\nu)\\
&-\frac{\pi}{2}\frac{\partial}{\partial A}\Bigg(
\frac{\Gamma(1+A)}{\Gamma\left(1+\frac{A-\Delta}{2}\right)
\Gamma\left(1+\frac{A+\Delta}{2}\right)}\Bigg)\Bigg\vert_{A=\Delta-d}.  
\end{split}
\end{equation}
For $A=\Delta-d-\epsilon$
\begin{equation}
\frac{1}{\Gamma\left(1+\frac{A-\Delta}{2}\right)}=\frac{\epsilon\Gamma(d/2)}{2}
(-1)^{d/2}+O(\epsilon^2)  
\end{equation}
and so
\beq
\int_0^{\pi/2}{\rm d}u\,k_1(u)\left(\cos\frac{\Delta\pi}{2}-\cos u\Delta
\right)=\cos\frac{\Delta\pi}{2}\,g^\prime(\nu)
+\frac{\pi}{4}(-1)^{d/2}\Gamma(d/2)\frac{\Gamma(1+\nu)}
{\Gamma(1+\alpha)}.~~~~
\eeq
Using this in (\ref{fin0}) we see that  the final result for the bulk field at the middle point simplifies to
(\ref{finaleven}).

\section{Symmetries}
\label{appB}

\subsection{AdS isometry $\longrightarrow$ boundary conformal transformation}
Let us use coordinates $(\rho,x)$ for a point $Y$ in the AdS bulk and perform
an AdS isometry $Y\longrightarrow gY\sim(\bar\rho,\bar x)$. In the large $\rho$ limit,
we write
\begin{equation}
\bar\rho=\rho+\sigma(g,x)+o(\rho),\quad
\bar x=gx+o(\rho),  
\end{equation}
where $o(\rho)$ vanishes in the $\rho\to\infty$ limit and
$x\longrightarrow gx$ is the boundary conformal transformation.
For the derivatives we have
\begin{equation}
\frac{\partial\bar\rho}{\partial\rho}=1+o(\rho),\qquad
\frac{\partial\bar\rho}{\partial x^A}=\frac{\partial\sigma(g,x)}{\partial x^A}+o(\rho), \quad
\frac{\partial\bar x^A}{\partial\rho}=o(\rho),\quad
\frac{\partial\bar x^A}{\partial x^B}=\frac{\partial(gx)^A}{\partial x^B}+o(\rho),
\end{equation}
which gives
\begin{equation}
M_g^{\rm AdS}(Y)=M_g^{\rm bound}(x)+o(\rho), 
\end{equation}
where
\begin{equation}
M_g^{\rm AdS}(Y)= \det \left(\frac{\partial (gY)}{\partial Y}\right),
\qquad\quad M_g^{\rm bound}(x)=\det
\left(\frac{\partial \bar x}{\partial x}\right).
\end{equation}
The AdS line element squared and the line element at the boundary are given by 
\begin{equation}
({\rm d}s^2)^{\rm AdS}=R^2{\rm d}\rho^2-R^2\cosh^2\rho{\rm d}t^2  
+R^2\sinh^2\rho {\rm d}n^i{\rm d}n^i, \quad ({\rm d}s^2)^{\rm bound}=-{\rm d}t^2  +{\rm d}n^i{\rm d}n^i,
\end{equation}
respectively.
For large $\rho$, a relation between measure factors, square root of corresponding metric determinants, is given by
\begin{equation}
\gamma^{\rm AdS}(Y)=R^{d+1}\left(\frac{e^\rho}{2}\right)^d \gamma^{\rm bound}(x).  
\end{equation}
Since $M_g^{\rm AdS}(Y)\gamma^{\rm AdS}(gY)=\gamma^{\rm AdS}(Y) $ for an AdS isometry $g$,
in the $\rho\to\infty$ limit we obtain
\begin{equation}
H^d(g,x)=J^d(g,x)\frac{\gamma^{\rm bound}(gx)}{\gamma^{\rm bound}(x)},\quad 
H(g,x):=e^{-\sigma(g,x)}, \ J(g,x)  := [M_g^{\rm bound}(x)]^{1/d}.
\label{DD77}
\end{equation}
From the group composition property $J(gh,x)=J(g,hx)J(h,x)$ and a similar relation
\begin{equation}
H(gh,x)=H(g,hx)H(h,x),
\label{Hrelation}
\end{equation}
we have
\begin{equation}
J(g,x)=\frac{1}{J(g^{-1},gx)} , \qquad H(g,x)=\frac{1}{H(g^{-1},gx)}.
\label{DD88}
\end{equation}

If $g$ is a boundary isometry (shift of the time coordinate $t$, rotation of~$n^i$),
we have\footnote{From now on we drop the superscript 'bound'.}
$J^d(g,x)\gamma(gx)=\gamma(x)$, which leads to $H(g,x)= 1$.

\subsection{BDHM formula}
Let us reconsider the BDHM relation
\begin{equation}
O(x)=\lim_{\rho\to\infty}(\sinh\rho)^\Delta\Phi(\rho,x).  
\end{equation}
Using the Fock space transformation property of the bulk field, $U(g)\Phi(Y)U^\dagger(g)=\Phi(gY)$,
we find the transformation law
\begin{equation}
\begin{split}  
U(g)O(x)U^\dagger(g)&=\lim_{\rho\to\infty}(\sinh\rho)^\Delta\Phi(\bar\rho,\bar x)=
\lim_{\rho\to\infty}\left(\frac{\sinh\rho}{\sinh\bar\rho}\right)^\Delta
(\sinh\bar\rho)^\Delta\Phi(\bar\rho,\bar x)\\
&=e^{-\Delta\sigma(g,x)}O(gx)=H^\Delta(g,x)O(gx).
\end{split}  
\label{transO}
\end{equation}

\subsection{Definition and properties of $I(Y,x)$}

The AdS invariant $S(Y_1,Y)$ for two bulk points $Y_1$ and $Y$ are given by
\begin{equation}
S(Y_1,Y)=\cosh\rho_1\cosh\rho\cos(t_1-t)
-\sinh\rho_1\sinh\rho\, n_1^i n^i,  
\label{Sinvariant}
\end{equation}
which satisfies $S(gY_1,gY)=S(Y_1,Y)$.
Its bulk-boundary version is defined by\footnote{In our usual coordinates
$I(Y,x)=2[\cosh\rho\cos(t-\tilde t)-\sinh\rho n^i \tilde n^i]$.}  
\begin{equation}
I(Y_1,x)=\lim_{\rho\to\infty}4e^{-\rho}S(Y_1,Y)
=2[\cosh\rho_1\cos(t_1-t)
-\sinh\rho_1\,n_1^i n^i].    
\end{equation}
Its transformation property is as follows.
\begin{equation}
I(gY_1,gx)=\lim_{\bar\rho\to\infty}4e^{-\bar\rho}S(gY_1,gY)=\lim_{\bar\rho\to\infty}
4e^{-\bar\rho}S(Y_1,Y)=H(g,x)I(Y_1,x).
\label{DD13}
\end{equation}

\subsection{Definition and properties of $T(Y,x)$}
Let us recall that a bulk point $Y$ of AdS space and a boundary point $x$
can be connected with a past directed light-like geodesic if $\tilde t=T_1$ with $ T_1:=t-\omega$,
where $\omega:=\arccos[(\tanh\rho)\,  n\cdot {\tilde n}]$ and $0<\omega<\pi$.
Similarly, $Y$ and $x$ can be connected with
a future directed light-like geodesic if $\tilde t=T_2$ with $T_2:=t+\omega$.
Finally, $Y$ and $x$ can be connected with
a space-like geodesic if $T_1<\tilde t<T_2$.

The function $T(Y,x)$, defined by $T(Y,x)=\Theta(T_2-\tilde t)\Theta(\tilde t-T_1)$
is isometry invariant as $T(gY,gx)=T(Y,x)$,
since the lightlike/spacelike nature of a curve in AdS space is
isometry invariant.

\section{Proof of $\hat\Phi(g)\equiv 0$}
\label{appC}
\subsection{Transformation back}

The transformation $g$ appears in several places in the definition of $\hat\Phi(g)$ but we can
simplify its expression by doing some steps backwards. Starting from the relation
\begin{equation}
I^{\Delta-d}(g^{-1}Y_o,y)=I^{\Delta-d}(Y_o,gy)H^{d-\Delta}(g,y)=
I^{\Delta-d}(Y_o,gy)\frac{J^d(g,y)\gamma(gy)}{\gamma(y)}H^{-\Delta}(g,y),    
\end{equation}
we write
\beqa
\hat\Phi(g) &=&\int{\rm d}^dy\gamma(y)I^{\Delta-d}(Y_o,gy)\frac{J^d(g,y)\gamma(gy)}
{\gamma(y)}H^{-\Delta}(g,y)T(Y_o,gy)\ln[H(g,y)]O(y)~~~    \\     
&=&
\int{\rm d}^dx\gamma(x)I^{\Delta-d}(Y_o,x)T(Y_o,x)H^\Delta(g^{-1},x)
\ln[H(g,g^{-1}x)]O(g^{-1}x)\\
&=&-\int{\cal D}xI^{\Delta-d}(Y_o,x)T(Y_o,x)
\ln[H(g^{-1},x)]U^\dagger(g)O(x)U(g)\\
&=& -U^\dagger(g)\hat\Psi(g^{-1})U(g),
\eeqa
where
\begin{equation}
\hat\Psi(g):=\int{\cal D}xI^{\Delta-d}(Y_o,x)T(Y_o,x)
\ln[H(g,x)]O(x).  
\end{equation}
From the group property (\ref{Hrelation}) satisfied by $H(g,x)$ it follows that
\begin{equation}
\hat\Psi(hg)=\hat\Psi(g)+
\int{\cal D}xI^{\Delta-d}(Y_o,x)T(Y_o,x)
\ln[H(h,gx)]O(x).  
\end{equation}
Therefore if $h$ is a boundary isometry ({\it i.e.} $H(h,{}^\forall y)=1$) then
\begin{equation}
\hat\Psi(hg)=\hat\Psi(g).  
\label{isoh}
\end{equation}

\subsection{The representation $g=b\Xi E$}

In this subsection we will use the embedding coordinates for global AdS.
\beq
X^i&=&R\sinh\rho\, n^i\ (i=1,\dots,d),\qquad X^0=R\cosh\rho\cos t, \quad
X^D=-R\cosh\rho\sin t .~~~
\eeq
The embedding coordinates satisfy $-(X^0)^2-(X^D)^2+X^i X^i=-R^2 $
and transform linearly under the AdS isometry SO$(d,2)$. The coordinates of the
middle point ($ t=0,\rho=0$) are 
\begin{equation}
Y_o:\quad X^i =0 \  (i=1,\dots,d), \quad X^0=R,\quad X^D=0.
\end{equation}
For an arbitrary bulk point $Y$ we will find a group element $g=b\Xi E$ such
that $g^{-1}Y_o=Y$. In other words, we transform $Y$ to $Y_o$ in three steps:
$EY =Y_2$, $\Xi Y_2=Y_1$, $bY_1 = Y_o$.

The first step ($E$) is a constant shift of the $t$ coordinate (SO$(2)$ rotation
in the $(0,D)$ plane) that brings $t$ to zero. After this step we have
\begin{equation}
Y_2:\quad
X^i=R\sinh\rho\, n^i \  (i=1,\dots,d), \quad X^0=R\cosh\rho, \quad X^D=0.
\end{equation}
The next step ($\Xi$) is an SO$(d)$ rotation in the $(1,\dots,d)$ space
that rotates the unit vector $n^i$ so that it becomes parallel to the $d$ axis.
Then we have
\begin{equation}
Y_1:\quad
X^i=0 \  (i=1,\dots,d-1), \quad
X^d=R\sinh\rho, \quad
X^0=R\cosh\rho, \quad
X^D=0.
\end{equation}
The last step ($b$) is a boost in the $(0,d)$ plane by $\bar X^d=X^d\cosh\beta-X^0\sinh\beta$, 
$\bar X^0=X^0\cosh\beta-X^d\sinh\beta$,  which makes $\bar X^d=0$, $\bar X^0=R$ for $\beta=\rho$.

This representation is useful because both $E$ and $\Xi$ are boundary isometries and
by (\ref{isoh}) we have 
\begin{equation}
\hat\Psi(g^{-1})=\hat\Psi(E^{-1}\Xi^{-1}b^{-1})=\hat\Psi(b^{-1}).  
\end{equation}
Hence the problem is reduced to the calculation of $\hat\Psi(b^{-1})$, where
$b^{-1}$ is the AdS isometry
\begin{equation}
\bar X^i=X^i, \
\bar X^d=X^d\cosh\beta+X^0\sinh\beta,\
\bar X^0=X^0\cosh\beta+X^d\sinh\beta, \
\bar X^D=X^D.
\end{equation}
Introducing polar coordinates $\theta,\omega$
as $n^d=\cos\theta$, $n^i=\sin\theta m^i(\omega)$ ($i=1,\dots,d-1$),
where $\omega$ are $d-1$ dimensional polar angles and $m^i$ is a $d-1$
dimensional unit vector, the transformation is given by
\begin{equation}
\bar\omega=\omega,\qquad
\left\{
\begin{split}
\sinh\bar\rho\sin\bar\theta&=\sinh\rho\sin\theta\\
\sinh\bar\rho\cos\bar\theta&=\sinh\rho\cos\theta\cosh\beta+
\cosh\rho\cos t\sinh\beta\\
\cosh\bar\rho\cos\bar t&=\cosh\rho\cos t\cosh\beta+
\sinh\rho\cos\theta\sinh\beta\\
\cosh\bar\rho\sin\bar t&=\cosh\rho\sin t
\end{split}  
\right.
\end{equation}
Consistency of the first two transformations determines $\bar\rho$:
\begin{equation}
\sinh^2\bar\rho=\sinh^2\rho\sin^2\theta+(\sinh\rho\cos\theta\cosh\beta+
\cosh\rho\cos t\sinh\beta)^2.  
\end{equation}
(Consistency of the second pair of transformations can also be used to determine
$\bar\rho$, but this is equivalent.)
We obtain $\bar\rho=\rho+\sigma$ in the $\rho\to\infty$ limit
\begin{equation}
e^{2\sigma}
=[\cosh\beta+\sinh\beta\cos(t-\theta)]  [\cosh\beta+\sinh\beta\cos(t+\theta)] .
\end{equation}
Using the convergent power series
\begin{equation}
\ln(1+u)=\sum_{k=1}^\infty\frac{(-1)^{k-1}}{k}u^k,  
\end{equation}
we have
\begin{equation}
\ln[H(b^{-1},x)]=\ln e^{-\sigma}=-\ln[\cosh\beta]+\sum_{k=1}^\infty\frac{(-\tanh\beta)^k}{k}
E_k(t,\theta),
\end{equation}
where
\begin{equation}
E_k(t,\theta)=\frac{1}{2}\left[\cos^k(t+\theta)+\cos^k(t-\theta)\right].
\end{equation}
The first few coefficients are
\beqa
E_1(t,\theta)&=&\cos\theta\cos t, \quad
E_2(t,\theta)=\cos^2\theta\cos^2 t+\sin^2\theta\sin^2 t, \nn \\
E_3(t,\theta)&=&\cos^3\theta\cos^3 t+3\cos\theta\cos t\sin^2\theta\sin^2 t .
\eeqa
The final result is
\begin{equation}
\hat\Psi(g^{-1})=\sum_{k=1}^\infty\frac{(-\tanh\rho)^k}{k}\hat\Psi_k,
\end{equation}
where
\begin{equation}
\hat\Psi_k=\int{\cal D}xI^{\Delta-d}(Y_o,x)T(Y_o,x)E_k(t,\theta)O(x).  
\label{psik}
\end{equation}

\subsection{Calculation of the first few terms}

(\ref{calO1}) gives the expansion of the boundary field:
\begin{equation}
O(t,\Omega)=\sum_{n\ell\underline{m}}\left[
e^{-i\nu_{n\ell}t}{\cal B}_{n\ell\underline{m}}
+e^{i\nu_{n\ell}t}{\cal B}^\dagger_{n\ell\underline{m}}\right]
Y_{\ell\underline{m}}(\Omega),
\label{Orep}
\end{equation}
where $\nu_{n\ell}=\Delta+\ell+2n$, $n=0,1,\dots$, $\ell=0,1,\dots$ and
${\cal B}_{n\ell\underline{m}}$ is the rescaled Fock space operator
\begin{equation}
{\cal B}_{n\ell\underline{m}}=\sqrt{\frac{{\cal N}R}{2\nu_{n\ell}}}\,
\frac{P_n(1+\alpha)}{n!}{\cal N}_{n\ell} {\cal A}_{n\ell\underline{m}}.  
\end{equation}

Using the coordinate system $\Omega=(\theta,\omega)$ in the previous subsection and ${\rm d}\Omega=\sin^{2a}\theta
{\rm d}\theta{\rm d}\omega$ with $a=-1+d/2$, 
we can write the $d$ dimensional spherical harmonics in terms of $d-1$ dimensional ones and Gegenbauer polynomials as
\begin{equation}
Y_{\ell\underline{m}}(\Omega)=Y_{\ell\lambda\underline{\tilde m}}(\theta,\omega)=
K_{\ell\lambda\underline{\tilde m}}\,C^{a+\lambda}_{\ell-\lambda}(\cos\theta)
\sin^\lambda\theta \,Y_{\lambda\underline{\tilde m}}(\omega),
\end{equation}
where the multi-index $\underline{m}$ is decomposed as $\lambda\,\underline{\tilde m}$, 
$K_{\ell\lambda\underline{\tilde m}}$ are some normalization constants to
ensure orthonormality, and the orthogonality of Gegenbauer polynomial is given by
\begin{equation}
\int_0^\pi{\rm d}\theta\sin^{2a}\theta\,C^a_\ell(\cos\theta)
C^a_{\ell^\prime}(\cos\theta)=\mu_\ell\delta_{\ell \ell^\prime}.
\label{mul}\end{equation}
With this choice we have
\begin{equation}
O(t,\Omega)=\sum_{n\ell\lambda\underline{\tilde m}}
K_{\ell\lambda\underline{\tilde m}}\left[
e^{-i\nu_{n\ell}t}{\cal B}_{n\ell\lambda\underline{\tilde m}}
+e^{i\nu_{n\ell}t}{\cal B}^\dagger_{n\ell\lambda\underline{\tilde m}}\right]
C^{a+\lambda}_{\ell-\lambda}(\cos\theta)\sin^\lambda\theta\,
Y_{\lambda\underline{\tilde m}}(\omega).
\end{equation}
Putting this expansion into (\ref{psik}) we obtain
\begin{equation}
\begin{split}
\hat\Psi_k=\sum_{n\ell\lambda\underline{\tilde m}}
&K_{\ell\lambda\underline{\tilde m}}
\int_{-\pi/2}^{\pi/2}{\rm d}t(2\cos t)^{\Delta-d}\left[
e^{-i\nu_{n\ell}t}{\cal B}_{n\ell\lambda\underline{\tilde m}}
+e^{i\nu_{n\ell}t}{\cal B}^\dagger_{n\ell\lambda\underline{\tilde m}}\right]\\
&\times \int_0^\pi{\rm d}\theta\sin^{2a}\theta
\,C^{a+\lambda}_{\ell-\lambda}(\cos\theta)\sin^\lambda\theta E_k(t,\theta)\,
\int{\rm d}\omega Y_{\lambda\underline{\tilde m}}(\omega).
\end{split}
\end{equation}
The last integral simply gives $\sqrt{\Omega_{d-1}}\delta_{\lambda 0}
\delta_{\underline{\tilde m}\underline{\tilde 0}}$, where $\Omega_{d-1}$ is the $d-1$
dimensional volume element. 
The formula further simplifies with $\hat b_{n\ell}=K_{\ell 0\underline{\tilde 0}}\sqrt{\Omega_{d-1}}
{\cal B}_{n\ell 0\underline{\tilde 0}}$ as
\begin{equation}
\hat\Psi_k=\sum_{n\ell}
\int_{-\pi/2}^{\pi/2}{\rm d}t(2\cos t)^{\Delta-d}\left[
e^{-i\nu_{n\ell}t}\hat b_{n\ell}
+e^{i\nu_{n\ell}t}\hat b^\dagger_{n\ell}\right]
\int_0^\pi{\rm d}\theta\sin^{2a}\theta
\,C^a_\ell(\cos\theta)\, E_k(t,\theta).
\end{equation}

We then write $\hat\Psi_k = \hat \Psi^{\rm ann}_k + \left(\hat\Psi^{\rm ann}_k\right)^\dagger$ with 
\beqa 
\hat\Psi^{\rm ann}_k
&=&\sum_{\ell=0}^k\oint\frac{{\rm d}z}{z}(-z)^{d/2}
(1-z)^{\Delta-d}\,B_\ell(z)P_{k\ell}(z),
\label{BlPkl}
\eeqa
where $z= - e^{-2it}$,  and
\beqa
\mu_\ell e^{i \ell t} P_{k\ell}(z):= \int_0^\pi {\rm d}\theta\sin^{2a}\theta
\,C^a_\ell(\cos\theta)\, E_k(t,\theta),
\quad
B_\ell(z):=\frac{ \mu_\ell}{2i}\sum_{n=0}^\infty \hat b_{n\ell}(-z)^n.  
\label{eq:Pkl}
\eeqa
Note that $P_{k\ell}(z)=0$  if $k+\ell$ is odd.

Let us calculate the first few terms. For this we need 
the Gegenbauer polynomials and their inverse relations\footnote{From here on we drop the superscript $a$.} such as
\beq
C_0(w)&=&1,\quad C_1(w)=2a w,\quad C_2(w)=a[2(a+1) w^2-1], \\
w&=&\frac{1}{2a}C_1(w),\quad
w^2=\frac{1}{2(a+1)}\left[\frac{1}{a}C_2(w)+1\right],
\eeq
to obtain
\begin{equation}
P_{11}(z)=\frac{1-z}{4a},\quad
P_{22}(z)=\frac{1+z^2}{4a(a+1)},\quad
P_{20}(z)=\frac{1}{2}+\frac{a}{4(a+1)}\left(z+\frac{1}{z}\right).  
\end{equation}
Using these results we see that 
\begin{equation}
\hat\Psi^{\rm ann}_1=\frac{1}{4a}\oint\frac{{\rm d}z}{z}(-z)^{d/2}
(1-z)^{\Delta-d}B_1(z)(1-z)=0, 
\end{equation}
because the integrand is analytic inside the unit circle.
Similarly
\begin{equation}
\hat\Psi^{\rm ann}_2=\oint\frac{{\rm d}z}{z}(-z)^{d/2}
(1-z)^{\Delta-d}\left\{B_2(z)\frac{1+z^2}{4a(a+1)}+B_0(z)
\left[\frac{1}{2}+\frac{a}{4(a+1)}\left(z+\frac{1}{z}\right)\right]
\right\}=0,
\end{equation}
since the $1/z$ pole coming from $P_{20}$ is compensated by the factor
$\frac{1}{z}(-z)^{d/2}$ for even integer $d > 2$.
In general, $P_{k\ell}(z)$ cannot be more singular
than $z^{-a}$, as we will see.

\subsection{General proof}
Using
\begin{equation}
E_k(t,\theta)
=\frac{1}{2^k}\sum_{r=0}^k\binom{k}{r}e^{i(2r-k)t}\cos(2r-k)\theta,  
\end{equation}
$P_{k\ell}(z)$ in \eqref{eq:Pkl} is evaluated as
\beqa
P_{k\ell}(z)&=&\frac{1}{\mu_\ell}\frac{1}{2^k}\sum_{r=0}^k\binom{k}{r}
e^{i(2r-k-\ell)t}\int_0^\pi{\rm d}\theta\cos(2r-k)\theta\sin^{2a}\theta
C^a_\ell(\cos\theta) \nn \\
&=&  \frac{1}{\mu_\ell}\frac{1}{2^k}\sum_{r=0}^k\binom{k}{r}
(-z)^{{\ell+k\over 2}-r} I_\ell^{(2r-k)},
\eeqa
where
\begin{equation}
I^{(n)}_\ell
=\frac{1}{2}\int_{-\pi}^\pi{\rm d}\theta e^{in\theta}\sin^{2a}\theta
C^a_\ell(\cos\theta).
\end{equation}
Since $\sin^{2a}\theta C^{a}_\ell(\cos\theta)$ can be written as a
Laurent polynomial in $e^{i\theta}$ of maximal degree $2a+\ell$, 
\begin{equation}
I^{(n)}_\ell=0\quad{\rm for}\quad n>2a+\ell,  
\end{equation}
which implies that 
$P_{k\ell}(z)$ cannot be more singular
than $z^{-a}$ as 
\begin{equation}
P_{k\ell}(z) = \frac{1}{\mu_\ell}\frac{1}{2^k}\sum_{n=-a}^{j}
\Theta(k-\ell+2n+1)
\binom{k}{j-n}
(-z)^{n} I_\ell^{(\ell-2n)}, \quad j:= \frac{k+\ell}{2}.
\end{equation}

\section{Recursion relations}
\label{appD}
\subsection{Recursion relation for $J_1(\nu,\omega)$}
Using the simple identity
\begin{equation}
\frac{{\rm d}}{{\rm d}u}(\cos u+u\sin u)=u\,\cos u
\end{equation}
we can perform a partial integration in the definition of $J_1(\nu,\omega)$
as follows. Here we require that $\nu>1$ so that all subsequent manipulations
are well-defined.
\beqa
J_1(\nu,\omega)&=&{1\over \Gamma(\nu+1)}
\int_0^\omega{\rm d}u(u\cos u-u\cos\omega)(\cos u-\cos\omega)^{\nu-1}\nn \\
&=&-{\cos\omega\over \nu} J_1(\nu-1,\omega) -{(\cos u-\cos\omega)^{\nu-1}\over \Gamma(\nu+1)}\nn\\
&+&{\nu-1\over \Gamma(\nu+1)}\int_0^\omega{\rm d}u(\cos u+u\sin u)(\cos u-\cos\omega)^{\nu-2}\sin u\nn\\
&=&-{\cos\omega\over \nu} J_1(\nu-1,\omega)-{(1-\cos\omega)^{\nu-1}\over \Gamma(\nu+1)}+{\nu-1\over \Gamma(\nu+1)}\int_0^\omega{\rm d}u
\big\{u\sin^2 u(\cos u-\cos\omega)^{\nu-2}\big\}\nn\\
&+&{\nu-1\over\Gamma(\nu+1)}\int_0^\omega{\rm d}u\,\cos u(\cos u-\cos\omega)^{\nu-2}\sin u.   
\eeqa
In the next to last line the integral, using $\sin^2 u=1-\cos^2 u$, can be
represented as
\begin{equation}
\Gamma(\nu-1)\sin^2\omega J_1(\nu-2,\omega)-\Gamma(\nu+1)J_1(\nu,\omega)-2\cos\omega \Gamma(\nu) J_1(\nu-1,\omega)
\end{equation}
The integral in the last line can be done explicitly and gives
\begin{equation}
\frac{1}{\nu}(1-\cos\omega)^\nu+\frac{\cos\omega}{\nu-1}(1-\cos\omega)^{\nu-1}.  
\end{equation}
Putting everything together, after some simplifications we get
\begin{equation}
\sin^2\omega J_1(\nu-2,\omega)=\nu^2 J_1(\nu,\omega)+(2\nu-1)\cos\omega
J_1(\nu-1,\omega)+\frac{1}{\Gamma(\nu+1)}(1-\cos\omega)^\nu.  
\end{equation}
Making the shift $\nu\rightarrow\nu+2$ (so that the result is now valid
for $\nu>-1$), we finally arrive at the recursion \eqref{eq:recursion}.

\subsection{Recursion relation for $P_1(\nu,\omega)$}
Using integration by parts we obtain
\beqa
P_1(\nu,\omega)&=&{1\over \Gamma(\nu+1)}\int_0^\omega{\rm d}u\,(\cos u-\cos\omega)
(\cos u-\cos\omega)^{\nu-1}\ln(\cos u-\cos\omega)\nn \\
&=&-{\cos\omega\over \nu}  P_1(\nu-1,\omega)
+\int_0^\omega{\rm d}u\,{\sin^2 u\over \Gamma(\nu+1)}\Big\{(\nu-1)(\cos u-\cos\omega)^{\nu-2}
\ln(\cos u-\cos\omega)\nn \\
&&+(\cos u-\cos\omega)^{\nu-2}\Big\},
\eeqa
and use the identity
\begin{equation}
\sin^2 u=\sin^2\omega-2\cos\omega(\cos u-\cos\omega)-(\cos u-\cos\omega)^2
\end{equation}
to get
\beqa
P_1(\nu,\omega)&=&-{\cos\omega\over\nu} P_1(\nu-1,\omega)+\sin^2\omega\left[{P_1(\nu-2,\omega)\over\nu}
+{K_1(\nu-2,\omega)\over \nu(\nu-1)}\right] \nn \\
&-&{2\cos\omega\over \nu}\left[(\nu-1)P_1(\nu-1,\omega)+K_1(\nu-1,\omega)\right]-(\nu-1)P_1(\nu,\omega)
-K_1(\nu,\omega).~~~~~~~
\eeqa
We here again assume $\nu>1$.
After some rearrangements  we obtain
\begin{equation}
\begin{split}
\sin^2\omega\, P_1(\nu-2,\omega)
&=(2\nu-1)\cos\omega P_1(\nu-1,\omega)+\nu^2 P_1(\nu,\omega)\\
&+2\cos\omega  K_1(\nu-1,\omega)+\nu K_1(\nu,\omega)-
\frac{\sin^2\omega}{\nu-1} K_1(\nu-2,\omega).
\end{split}
\end{equation}
Finally we make the shift $\nu\to \nu+2$ (so that the recursion is valid for $\nu>-1$) to obtain \eqref{eq:recursionP}.

\section{AdS Green's functions}
\label{space1}

In this appendix we construct the space-like Green's function in AdS space,
which will be useful (see appendix \ref{space2}) in an alternative method
\cite{Hamilton:2006az,Heemskerk:2012mn} of the bulk reconstruction. Here, for
simplicity, we will restrict our considerations to the range $\Delta>d-2$ only.

A Green's function of the massive scalar wave equation satisfies
\begin{equation}
({\cal D}-m^2){\cal G}(Y,Y^\prime)=\frac{1}{\sqrt{|g|}}\delta(Y,Y^\prime),
\quad {\cal D}:=\frac{1}{\sqrt{|g|}}\partial_\alpha(\sqrt{|g|}g^{\alpha\beta}\partial_\beta),
\end{equation}
where the Laplacian ${\cal D}$ acts on $Y$, 
and the mass is parametrized as $m^2=\Delta(\Delta-d)$ (from now on we are
using units where the AdS radius is unity).
The metric $g_{\alpha\beta}$ (and its determinant $g$ and inverse
$g^{\alpha\beta}$) in Lorentzian\ AdS
 is encoded by the line element ${\rm d}s^2$ as
\begin{equation}
  {\rm d}s^2=
({\rm d}\rho)^2-\cosh^2\rho\,({\rm d}t)^2+\sinh^2\rho\,{\rm d}n^i{\rm d}n^i
\end{equation}
in global\ coordinates, or 
\begin{equation}
 {\rm d}s^2=
-(1+y^2)({\rm d}t)^2+\big(\delta_{ij}-
\frac{y^iy^j}{1+y^2}\big){\rm d}y^i{\rm d}y^j
\end{equation}
in  flat\ coordinates, where $y^i=n^i\sinh\rho $ and $y:=\sqrt{y^i y^i}=\sinh\rho$.

In the Green's function method the AdS invariant (\ref{Sinvariant})
\begin{equation}
\sigma(Y,Y^\prime)=\cos(t-t^\prime)\cosh\rho\,\cosh\rho^\prime
- \underline{n}\cdot\underline{n}^\prime  \sinh\rho\,\sinh\rho^\prime 
\end{equation}
will play an important role, where  $\underline{n}$, $\underline{n}^\prime$ are  $d$-dimensional vectors. 
If we are looking for a $\sigma$-dependent Green's function ${\cal G}(Y,Y^\prime)=g(\sigma(Y,Y^\prime))$,
then $g(\sigma)$ has to satisfy the differential equation  
\begin{equation}
(\sigma^2-1)g^{\prime\prime}(\sigma)+D\sigma g^\prime(\sigma)+\Delta(d-\Delta)
g(\sigma)=\frac{1}{\sqrt{|g|}}\delta(Y,Y^\prime).  
\label{gsigma}
\end{equation}
If $Y^\prime=Y_o$, we can take the
more general ansatz in flat coordinates  as ${\cal G}(Y,Y_o)={\cal H}(t,y)$, which should satisfy
\begin{equation}
-m^2{\cal H}(t,y)-\frac{1}{1+y^2}\partial_t^2{\cal H}(t,y)+\partial_i
\left[{y^i\over y}\big(1+y^2\big)\partial_y{\cal H}(t,y)\right]=\delta(t)
\delta(\underline{y}).
\end{equation}
The delta function normalization of this Green's function becomes more
transparent in terms of its Fourier transform
\begin{equation}
H(\omega,y)=\int_{-\infty}^\infty{\rm d}t\,{\rm e}^{i\omega t}\,{\cal H}(t,y).  
\end{equation}
We have to require
\begin{equation}
\begin{split}
y>&0:\ \ 
\Big[\frac{\omega^2}{1+y^2}+\Delta(d-\Delta)\Big]H(\omega,y)+\partial_i
\left[{y^i\over y}\big(1+y^2\big)\partial_y H(\omega,y)\right]=0,\\
y\to&0:\ \ 
H(\omega,y)\approx -\frac{y^{2-d}}{(d-2)\Omega_d}.
\end{split}  
\end{equation}

\subsection{Hypergeometric $\sigma$-dependent solutions}

A particular (properly normalized) solution of (\ref{gsigma}) is given
by the hypergeometric solution (see \cite{DHoker:2002nbb})
\begin{equation}
F(\sigma)=-\frac{\Gamma(\Delta)}{2^{\Delta+1}\pi^{d/2}\Gamma(1+\alpha)}
\sigma^{-\Delta} {}_2F_1\left(\frac{\Delta}{2},\frac{\Delta+1}{2};1+\alpha;
\frac{1}{\sigma^2}\right),  
\end{equation}
which is the scalar two-point correlation function in AdS space.
This solution is singular for $\sigma\to1$ and is properly normalized as
\begin{equation}
F(\sigma)\sim d_*x^{-a},\qquad a=\frac{d-1}{2},\qquad d_*=-\frac{\Gamma(D/2)}
{(D-2)(2\pi)^{D/2}}  
\end{equation}
for $x\to0$ with $\sigma=1+x$.
A solution of the homogeneous part of (\ref{gsigma}) can also be given
\cite{Aoki:2022lye} in terms of a hypergeometric function:
\begin{equation}
J(\sigma)=  
{}_2F_1\left(\frac{\Delta}{2},\frac{d-\Delta}{2};\frac{d+1}{2};1-\sigma^2\right).  
\end{equation}
This solution is regular at $\sigma=1$: $J(1)=1$.

Although the solution of the problem is completely given
\cite{Hamilton:2006az,Heemskerk:2012mn} (see also \cite{Bhattacharjee:2022ehq})
in terms of these two special functions, it is nevertheless more transparent if
we write the Green's function in an expanded form using the variable
$x=\sigma-1$. We take the ansatz
\begin{equation}
g(\sigma)=\psi(x)=\sum_{n=0}^\infty d_{n+q}x^{n+q} 
\label{xexp}
\end{equation}
and expand the homogeneous part of the equation written as
\begin{equation}
x(x+2)\psi^{\prime\prime}(x)+D(x+1)\psi^\prime(x)+\Delta(d-\Delta)\psi(x)=0.  
\label{xeq}
\end{equation}
The regular solution corresponds to the choice $q=0$ and we write
\begin{equation}
J(\sigma)=h(x)=\sum_{n=0}^\infty c_nx^n,  
\end{equation}
where the expansion coefficients are given recursively as
\begin{equation}
c_{n+1}=-\frac{(n+\Delta)(n+d-\Delta)}{(n+1)(2n+d+1)}\,
c_n\qquad n=0,1,\dots\qquad c_0=1.  
\end{equation}
The singular solution corresponds to $q=-a$ and proper normalization requires $d_{-a}=d_*$.
Higher coefficients are determined from the recursion
\begin{equation}
d_{n-a}(n-a+\Delta)(n-a+d-\Delta)+2d_{n-a+1}(n+1)(n+1-a)=0, \quad n=0,1,\dots \, .
\label{recd}  
\end{equation}
For even $d$ (when $a$ is half-integer), all higher $d_{n-a}$ coefficients are
obtained recursively from $d_{-a}=d_*$. On the other hand, for odd $d$ (when
$a$ is integer), we first determine the coefficients $d_{-a+1},\dots,d_{-1}$ from $d_{-a}$
using \eqref{recd}, which are all non-zero in the range $\Delta>d-2$.  Arriving at
$n=a-1$ in the recursion (\ref{recd}), we find a contradiction unless $\Delta=d-1$.
In this case it is consistent to put $d_n=0$, $n=0,1,\dots$ .

For generic $\Delta\not=d-1$ and odd $d$, there is no singular solution within
the ansatz (\ref{xexp}). 
We therefore take a different ansatz, 
\begin{equation}
g(\sigma)=\tilde\psi(x)=\psi(x)+c\ln x\, h(x),  
\end{equation}
which satisfies (\ref{xeq}) with the coefficients $d_{-a},\dots,d_{-1}$ as before.
Then the $n=a-1$ equation leads to 
\begin{equation}
c=\frac{d_{-1}}{d-1}(\Delta-1)(\Delta-d+1),  
\end{equation}
and the higher coefficients can be calculated from the recursion
\begin{equation}
\begin{split}
d_{n+1}=-\frac{1}{(n+1)(2n+d+1)}&\Big\{(n+\Delta)(n+d-\Delta)d_n\\
&+c\big[c_n(2n+d)+c_{n+1}(4n+d+3)\big]\Big\},\quad n=0,1,\dots
\end{split}  
\end{equation}
(By convention) we fix $d_0=0$ and the general $\sigma$-dependent solution is
then given by 
\begin{equation}
g(\sigma)+pJ(\sigma),  
\end{equation}
where $p$ is an arbitrary constant.

\subsection{Feynman propagator}

By analogy to the Minkowski case, we define the Feynman propagator in Lorentzian
AdS space as
\begin{equation}
G(\sigma)={\rm Re}\,\big\{ig(\sigma+i\epsilon)\big\}.
\end{equation}

\subsubsection{Odd $d$}

By using
\begin{equation}
\begin{split}
{\rm Re}\,\{i(x+i\epsilon)^n\}&=0,\qquad n\geq0,\\
{\rm Re}\,\Big\{\frac{i}{x+i\epsilon}\Big\}&=\pi\delta(x),\quad
{\rm Re}\,\Big\{\frac{i}{(x+i\epsilon)^{k+1}}\Big\}=\frac{(-1)^k\pi}{k!}
\delta^{(k)}(x),
\end{split}    
\end{equation}
we can write a contribution to the Feynman propagator coming from $\psi(x)$ as
\begin{equation}
G_{\rm s}(x)=\sum_{k=0}^{a-1}f_k\delta^{(k)}(x),\qquad
f_k=\frac{(-1)^k\pi}{k!}d_{-(k+1)},\quad k=0,\dots,a-1,  
\end{equation}
where contributions from non-singular terms vanish. 
For later use we rewrite the recursion relations in terms of the $f_k$
coefficients as
\begin{equation}
(k+1-\Delta)(k+1+\Delta-d)f_k+(d-2k-1)f_{k-1}=0,\qquad k=1,\dots,a-1,  
\label{recf}
\end{equation}
\begin{equation}
f_{a-1}=\frac{(-1)^a\sqrt{\pi}}{\pi^{d/2}2^{\frac{d+3}{2}}},\qquad
\pi c=\frac{(\Delta-1)(\Delta-d+1)}{d-1}\,f_0.    
\label{fam1}
\end{equation}
Using the relation
\begin{equation}
\ln(x+i\epsilon)\,h(x+i\epsilon)=\left\{
\begin{matrix}
(i\pi+\ln|x|)h(x)\quad&x<0\\
\ln x\, h(x)&x>0\end{matrix}\right.  ,
\end{equation}
we have
\begin{equation}
{\rm Re}\,\big[i\ln(x+i\epsilon)h(x+i\epsilon)\big]=-\pi h(x)\Theta(-x).
\end{equation}
Thus the full Feynman propagator (for odd $d$ in the Lorentzian AdS) becomes
\begin{equation}
G(\sigma)=G_{\rm s}(x)-c\pi h(x)\Theta(-x).  
\end{equation}
In the next section we will need the space-like Green's function, constructed as
\begin{equation}
\overline{g} (Y, Y^\prime)
=G(\sigma)+\pi cJ(\sigma)=G_{\rm s}(x)+c\pi h(x)\Theta(x).
\label{barG}
\end{equation}
This Green's function vanishes in the time-like region ($\sigma<1$,  equivalently, $x<0$), since $G_s(x)=0$ for $x\not=0$.

\subsubsection{Even $d$}

In this case the expansion is in half-integer powers and the Feynman propagator
is
\begin{equation}
G(\sigma)={\rm Re}\,\{ig(\sigma+i\epsilon)\}=\left\{
\begin{matrix}
0\quad&x>0\\
-\sqrt{|x|}\sum_{n=0}^\infty d_{n-a}x^{n-d/2}&x<0\end{matrix}\right.  
\end{equation}
Since the half-integer powers cannot be cancelled by adding a term of the
form $pJ(\sigma)$ we conclude that there is no $\sigma$-dependent space-like
Green's function for even $d$.

\section{Green's function method}
\label{space2}

The starting point here is the identity involving the Green's function
${\cal G}(Y,Y^\prime)$ and a massive free scalar field $\Phi(Y)$:
\begin{equation}
\begin{split}  
&\partial_\mu\big(\sqrt{-g}g^{\mu\nu}\partial_\nu {\cal G}\cdot \Phi-  
\sqrt{-g}g^{\mu\nu}\partial_\nu \Phi\cdot {\cal G}\big)=
\sqrt{-g}\big({\cal D}{\cal G}\cdot\Phi-{\cal D}\Phi\cdot{\cal G}\big)\\
&=\sqrt{-g}\big(({\cal D}-m^2){\cal G}\cdot\Phi
-({\cal D}-m^2)\Phi\cdot{\cal G}\big)=\delta(Y,Y^\prime)\Phi(Y).
\end{split}
\end{equation}
Integrating the above relation with respect to $Y$ and using Stokes' theorem, we obtain
\begin{equation}
\Phi(Y^\prime)=\int{\rm d}^DY\partial_\mu X^\mu=\oint {\rm d}n_\mu X^\mu,  
\quad X^\mu:=\sqrt{-g}g^{\mu\nu}(\partial_\nu{\cal G}\cdot \Phi-\partial_\nu\Phi\cdot {\cal G}),
\label{Stokes}
\end{equation}
where
the surface integral in (\ref{Stokes}) must include the bulk point
$Y^\prime$ in its interior. 
We now choose the space-like Green's function ${\cal G}=\overline{\cal G}$ for $Y^\prime=Y_o$, and furthermore
the surface is chosen to be a cylinder with symmetry axis parallel to the $t$
coordinate axis and radius $\rho=R$. The top and bottom bases of the cylinder
are at $t=\pm t_1$, where $\frac{\pi}{2}>t_1>t_o$ and $\cos t_o=\dfrac{1}{\cosh R}$.  
The two bases of the cylinder at $t=t_1$ and $t=-t_1$ do not contribute to the integral, since 
$\sigma<1$ uniformly
there and the space-like Green's function $\overline{\cal G}$ vanishes. 
Therefore we can write
\begin{equation}
\Phi(Y_o)=\cosh R(\sinh R)^{d-1}\int_{-t_1}^{t_1}{\rm d}t\int{\rm d}\Omega
[\partial_\rho\overline{\cal G}\cdot\Phi-
\partial_\rho\Phi\cdot\overline{\cal G}].
\end{equation}
Since $\overline{\cal G}(Y,Y_o)$ depends only on $t$ and $\rho$, the
angular integration leads to
\begin{equation}
\Phi(Y_o)=\cosh R(\sinh R)^{d-1}\int_{-t_1}^{t_1}{\rm d}t
[\partial_\rho\overline{\cal G}(t,R)\cdot D(t,R)-
\partial_\rho D(t,R)\cdot\overline{\cal G}(t,R)],
\end{equation}
where $\displaystyle D(t,\rho):=\int{\rm d}\Omega\, \Phi(t,\rho,\Omega)$ is the S-wave part of the scalar field.
For large $R$, it satisfies the BDHM relation $D(t,R)\approx (\sinh R)^{-\Delta}C(t)$,
where $C(t)$ is the S-wave component of the boundary conformal field.
Thus the bulk reconstruction formula simplifies to
\begin{equation}
\Phi(Y_o)=\lim_{R\to\infty}(\cosh R)^{-\nu}\int_{-t_1}^{t_1}{\rm d}t C(t)
[\partial_\rho\overline{\cal G}(t,R)+\Delta \overline{\cal G}(t,R)].
\end{equation}

Further simplification occurs for an odd $d$, because, as we have seen in
appendix \ref{space1},  $\overline{\cal G}$ only depends on
$\sigma$ in this case.  Therefore,
\begin{equation}
\partial_\rho \overline{G}(\sigma)=\cos t\sinh R\,\overline{G}^\prime(\sigma)
\approx \sigma\overline{G}^\prime(\sigma)  
\end{equation}
for large $R$, and we can write
\begin{equation}
\Phi(Y_o)=\lim_{R\to\infty}(\cosh R)^{-\nu}\int_{-t_1}^{t_1}{\rm d}t C(t)
[\sigma\overline{G}^\prime(\sigma)+\Delta \overline{G}(\sigma)].
\label{bulkbarG}
\end{equation}

Let us now separate the delta function  and theta function parts of the
representation in (\ref{bulkbarG}) as
\begin{equation}
\Phi(Y_o)=\lim_{R\to\infty}(\Phi_0+\Phi_1),  
\end{equation}
where
\begin{equation}
\Phi_0=(\cosh R)^{-\nu}\int_{-t_1}^{t_1}{\rm d}t C(t)
[\sigma G_{\rm s}^\prime(x)+\Delta G_{\rm s}(x)+\pi c\delta(x)]
\end{equation}
with $h(0)=J(1)=1$, and 
\begin{equation}
\Phi_1=\pi c(\cosh R)^{-\nu}\int_{-t_o}^{t_o}{\rm d}t C(t)
[\sigma J^\prime(\sigma)+\Delta J(\sigma)].
\end{equation}
The range of the $t$ integral  becomes $[-t_o,t_o]$ due to $\Theta(x)$, and we introduce $t_o=\frac{\pi}{2}-\varepsilon_o$, where
$\sin\varepsilon_o=1/\cosh R\to 0$ for large $R$.

For technical reasons, we will now divide $\Phi_1$ into two parts, $\Phi_1=\Phi_{1a}+\Phi_{1b}$,
where
\begin{equation}
\Phi_{1a}=\pi c(\cosh R)^{-\nu}\int_{-\pi/2+\varepsilon}^{\pi/2-\varepsilon}{\rm d}t
C(t)[\sigma J^\prime(\sigma)+\Delta J(\sigma)],
\end{equation}
and
\begin{equation}
\begin{split}  
\Phi_{1b}=\pi c(\cosh R)^{-\nu}\Big\{\int_{\pi/2-\varepsilon}^{\pi/2-\varepsilon_o}
{\rm d}t &C(t) [\sigma J^\prime(\sigma)+\Delta J(\sigma)]\\
&+\int_{-\pi/2+\varepsilon_o}^{-\pi/2+\varepsilon}{\rm d}t C(t)
[\sigma J^\prime(\sigma)+\Delta J(\sigma)]\Big\}.
\end{split}
\end{equation}
Here $\varepsilon$ is a small but fixed parameter, while $\varepsilon_o$
tends to zero as $R\to\infty$. We will let $\varepsilon\to0$ at the end of
the calculation.

\subsection{Calculation of $\Phi_{1a}$}

As discussed in appendix \ref{useful}, $J(\sigma)$ has a power-law behaviour
for large $\sigma$  if  we restrict our considerations to $\Delta>d/2$, which is relevant only for $d=3$ with an odd $d$ since $\Delta > d-2$. 
Using the asymptotics (see appendix
\ref{useful}) 
\begin{equation}
\sigma J^\prime(\sigma)+\Delta J(\sigma)\approx(\nu+\Delta)G_o\sigma^\nu
\end{equation}
for $\sigma>\sin\varepsilon\cosh R\to\infty$,  
we have
\begin{equation}
\Phi_{1a}\approx \pi c G_o(2\alpha)(\cosh R)^{-\nu}
\int_{-\pi/2+\varepsilon}^{\pi/2-\varepsilon}{\rm d}t\, C(t)(\cos t\cosh R)^\nu =2^\nu\xi \int_{-\pi/2+\varepsilon}^{\pi/2-\varepsilon}{\rm d}t\, C(t)(\cos t)^\nu.
\end{equation}
In the following we will continue the calculation for the cases: \framebox{$\cal A$} \  $ \nu>-1$, \
\framebox{$\cal B$} \  $ \nu=-1$, \ \framebox{$\cal C$} \  $-1> \nu>-2$,
separately. For the simplest case, 
\begin{equation}
\framebox{$\cal A$}\quad \Phi_{1a}\approx \xi\int_{-\pi/2}^{\pi/2}{\rm d}t\,
C(t)(2\cos t)^\nu,  
\end{equation}
since the integral is convergent for $\varepsilon\to0$ in this range, while
\framebox{$\cal B$}\ $\Phi_{1a}=0$,
since $\xi$ vanishes in this spacial case. In the most complicated case
we use 
a partial integration and obtain
\begin{equation}
\begin{split}  
\framebox{$\cal C$}\quad \Phi_{1a}&=-2^\nu\xi
\int_{-\pi/2+\varepsilon}^{\pi/2-\varepsilon}{\rm d}t\,\dot{C}(t)g_1(t)+
2^\nu\xi\big[C(\pi/2-\varepsilon)+C(-\pi/2+\varepsilon)\big]
g_1(\pi/2-\varepsilon)\\
&\approx -2^\nu\xi \int_{-\pi/2}^{\pi/2}{\rm d}t\,\dot{C}(t)g_1(t)+
2^\nu\xi C_+(\varepsilon)\int_{\varepsilon}^{\pi/2}{\rm d}u(\sin u)^\nu,
\end{split}
\end{equation}
where $g_1(t)$ is the primitive function
\begin{equation}
g_1(t)=\int_0^t{\rm d}u(\cos u)^\nu.
\end{equation}
The last integral can be evaluated as (see appendix \ref{useful})
\begin{equation}
\int_\varepsilon^{\pi/2}{\rm d}u(\sin u)^\nu=  
\int_\varepsilon^{\pi/2}{\rm d}u [(\sin u)^\nu-u^\nu]+\frac{1}{\nu+1}\left(
\frac{\pi}{2}\right)^{\nu+1}-\frac{\varepsilon^{\nu+1}}{\nu+1}\approx
\tilde g_1-\frac{\varepsilon^{\nu+1}}{\nu+1}.
\end{equation}
Putting elements of this calculation together, we finally obtain
\begin{equation}
\Phi_{1a}=\xi\int_{\rm (sub)}{\rm d}t \, (2\cos t)^\nu\, C(t)+\frac{\eta}{2\Omega_d}
C_+(\varepsilon)-2^\nu\xi C_+(\varepsilon)\frac{\varepsilon^{\nu+1}}{\nu+1}, 
\end{equation}
where the first term is written in terms of the subtracted integral, defined in \eqref{eq:int_subt}, by
reversing the partial integration.

\subsection{Calculation of $\Phi_{1b}$}
As a first step, we simplify $\Phi_{1b}$ as follows:
\begin{equation}
\Phi_{1b}\approx \pi c C_+(0)(\cosh R)^{-\nu}\int_1^M\frac{{\rm d}\sigma}
{\sqrt{(\cosh R)^2-\sigma^2}}\big[\sigma J^\prime(\sigma)+\Delta J(\sigma)\big],
\end{equation}
where the upper limit $M=\sin\varepsilon\cosh R$ is  large. 
Next using
\begin{equation}
\frac{1}{\sqrt{1-\frac{\sigma^2}{(\cosh R)^2}}}-1\leq\frac{1}{\cos\varepsilon}-1
={\rm O}(\varepsilon^2)    
\end{equation}
we can further approximate $\Phi_{1b}$ as
\begin{equation}
\Phi_{1b}\approx \pi c C_+(\varepsilon)(\cosh R)^{-(\nu+1)}
{\cal F}(\sin\varepsilon\cosh R),  \quad {\cal F}(M):=\int_1^M{\rm d}\sigma\big[\sigma J^\prime(\sigma)
+\Delta J(\sigma)\big].
\end{equation}
Using the asymptotic formula $J(\sigma)\approx G_o\sigma^\nu$ again, we obtain
\begin{equation}
{\cal F}^\prime(M)\approx (\nu+\Delta)G_o M^\nu   \longrightarrow {\cal F}(M)\approx \frac{2\alpha G_o}{\nu+1} M^{\nu+1}+{\rm const.} \ ,
\end{equation}
with which  $\Phi_{1b}$ is evaluated case by case as before.
In the first case the constant term
is subleading and we obtain
\begin{equation}
\framebox{$\cal A$}\quad \Phi_{1b}\approx \frac{2^\nu\xi}{\nu+1}
(\sin\varepsilon)^{\nu+1}C_+(\varepsilon)\rightarrow0 ,
\end{equation}
while  \framebox{$\cal B$}\quad $\Phi_{1b}\approx 0\ \ (c=0)$
for the next case.
For the last case, the constant term dominates and the $M^{\nu+1}$
contribution is subleading, so that
we obtain
\begin{equation}
\framebox{$\cal C$}\quad \Phi_{1b}\approx \frac{2^\nu\xi}{\nu+1}
(\sin\varepsilon)^{\nu+1}C_+(0)+\pi c C_+(0)(\cosh R)^{-(\nu+1)}{\cal F}(\infty).
\end{equation}
where the results in appendix \ref{useful} gives
\begin{equation}
{\cal F}(\infty)=-1+(\Delta-1)\int_1^\infty{\rm d}\sigma J(\sigma)=-1-
\frac{d-1}{\nu+1}.  
\end{equation}

We can now add up the contributions $\Phi_{1a}$ and $\Phi_{1b}$ and find
\begin{equation}
\framebox{$\cal A$}\quad \Phi_1=\xi\int_{-\pi/2}^{\pi/2}{\rm d}t
C(t)(2\cos t)^\nu,  
\end{equation}
\begin{equation}
\framebox{$\cal B$}\quad \Phi_1=0,  
\end{equation}
\begin{equation}
\framebox{$\cal C$}\quad \Phi_1=\xi\int_{\rm (sub)}{\rm d}t
C(t)(2\cos t)^\nu +\frac{\eta}{2\Omega_d}C_+(0)+\pi c C_+(0){\cal F}(\infty)
(\cosh R)^{-(\nu+1)}.
\end{equation}

\subsection{Calculation of $\Phi_0$}
Using the delta function identity $x\delta^{(k)}(x)=-k\delta^{(k-1)}(x)$,
$\Phi_0$ can be rewritten as
\begin{equation}
\begin{split}
\Phi_0=(\cosh R)^{-\nu}\int_{-t_1}^{t_1}{\rm d}t\, C(t)&\big\{f_{a-1}\delta^{(a)}(x)
+\sum_{k=1}^{a-1}[f_{k-1}+(\Delta-k-1)f_k]\delta^{(k)}(x)\\
&+[(\Delta-1)f_0+\pi c]\delta(x)\big\}.   
\end{split}    
\label{Phi0v}
\end{equation}
Using the relations
\begin{equation}
\int_0^{t_1}{\rm d}t C(t)\delta^{(k)}(x)=(-1)^k\left[\left(\frac{{\rm d}}  
{{\rm d}\sigma}\right)^k\,\frac{C(t)}{\sqrt{(\cosh R)^2-\sigma^2}}\right]
\Bigg|_{t=t_o,\sigma=1},\quad \frac{{\rm d}t}{{\rm d}\sigma}=-\frac{1}
{\sqrt{(\cosh R)^2-\sigma^2}}
\label{deltap}
\end{equation}
and
\begin{equation}
\int_{-t_1}^0{\rm d}t C(t)\delta^{(k)}(x)=(-1)^k\left[\left(\frac{{\rm d}}  
{{\rm d}\sigma}\right)^k\,\frac{C(t)}{\sqrt{(\cosh R)^2-\sigma^2}}\right]
\Bigg|_{t=-t_o,\sigma=1},\quad \frac{{\rm d}t}{{\rm d}\sigma}=\frac{1}
{\sqrt{(\cosh R)^2-\sigma^2}}
\label{deltam}
\end{equation}
we can evaluate (\ref{Phi0v}) term by term. Starting with $k=0$, the sum of the
delta function integrals (\ref{deltap}) and (\ref{deltam}) give
\begin{equation}
(k=0)\quad \frac{C(t_o)+C(-t_o)}{\sinh R},
\label{leadk0}
\end{equation}
\begin{equation}
(k=1)\quad \frac{C^\prime(t_o)-C^\prime(-t_o)}{\sinh^2 R}
-\frac{C(t_o)+C(-t_o)}{\sinh^3 R},
\end{equation}
\begin{equation}
(k=2)\quad \frac{C^{\prime\prime}(t_o)+C^{\prime\prime}(-t_o)}{\sinh^3 R}
-\frac{3[C^\prime(t_o)-C^\prime(-t_o)]}{\sinh^4 R}
+\frac{C(t_o)+C(-t_o)}{\sinh^3 R}+ \frac{3[C(t_o)+C(-t_o)]}{\sinh^5 R},
\end{equation}
and so on. We see that the leading contribution is given by (\ref{leadk0})
and all higher contributions are subleading (of order $(\sinh R)^{-(k+1)}$).
After this simplification, we find
\begin{equation}
\Phi_0\approx(\cosh R)^{-(\nu+1)}C_+(0)[(\Delta-1)f_0+\pi c],
\end{equation}
which, using the results in appendix \ref{useful}, gives
\begin{equation}
\framebox{$\cal A$}\quad \lim_{R\to\infty}\Phi_0=0,  
\end{equation}
\begin{equation}
\framebox{$\cal B$}\quad c=0\quad (\Delta-1)f_0=\frac{(-1)^a}{2\Omega_d}\quad
\Phi_0=\frac{(-1)^a}{2\Omega_d}C_+(0),  
\end{equation}
\begin{equation}
\framebox{$\cal C$}\quad \Phi_0\approx(\cosh R)^{-(\nu+1)}C_+(0) c\pi\left\{
1+\frac{d-1}{\nu+1}\right\}.
\end{equation}

The final result of the bulk reconstruction by the Green's function method is
given as
\begin{equation}
\begin{split}
\framebox{$\cal A$}\quad \Phi(Y_o)&=\xi\int_{-\pi/2}^{\pi/2}{\rm d}t\, C(t)
(2\cos t)^\nu = \eqref{eq:odd_HKLL} ,\\
\framebox{$\cal B$}\quad \Phi(Y_o)&=\frac{(-1)^a}{2\Omega_d}C_+(0) = \eqref{eq:ours_special} \ [\mbox{for $\ell=0$}],\\
\framebox{$\cal C$}\quad \Phi(Y_o)&=\xi\int_{\rm (sub)}{\rm d}t\, C(t)
(2\cos t)^\nu +\frac{\eta}{2\Omega_d}C_+(0) =  \eqref{eq:odd_ours}  .
\end{split}    
\end{equation}
These Green's function results are in complete agreement with those in the
main text obtained by different methods, as shown in the last equalities.

\section{Useful relations}
\label{useful}

In this appendix we list some results which will be used in the Green's function
method.
\begin{itemize}
\item  
From the asymptotics of hypergeometric functions we see that for
large argument $\sigma$
\begin{equation}
J(\sigma)\approx G_o\sigma^\nu,
\label{as}
\end{equation}
which is valid for $\Delta>d/2$ only.
Since we consider
the range $\Delta>d-2$ in this paper, this restriction is only relevant for
$d=3$. The coefficient in (\ref{as}) is given by
\begin{equation}
G_o=2^{\Delta-1}\frac{\Gamma(D/2)\Gamma(\alpha)}{\sqrt{\pi}\,\Gamma(\Delta)}.
\end{equation}

\item  
The integral of the hypergeometric solution $J(\sigma)$ can be calculated
with the help of the following two hypergeometric identities:
\begin{equation}
(1-z)^{a+b-c}{}_2F_1(a,b;c;z)={}_2F_1(c-a,c-b;c;z),  
\end{equation}
\begin{equation}
{}_2F_1(a,b;c;z)=\frac{c-1}{(a-1)(b-1)}\frac{{\rm d}}{{\rm d}z}{}_2
F_1(a-1,b-1;c-1;z).  
\end{equation}
With the substitution $\sigma=\sqrt{1+z}$ the integral is calculated to be
\begin{equation}
\int_1^\infty{\rm d}\sigma J(\sigma)=\frac{1}{2}\int_0^\infty\frac{{\rm d}z}
{\sqrt{1+z}}{}_2F_1\left(\frac{\Delta}{2},\frac{d-\Delta}{2};\frac{d+1}{2};  
-z\right)=-\frac{d-1}{(\nu+1)(\Delta-1)}.  
\end{equation}

\item  
For the parameter range $\nu>-2$, $\nu\not=-1$ the constant $\tilde g_1$
given below is well-defined and is given by
\begin{equation}
\tilde g_1=\int_0^{\pi/2}{\rm d}u[(\sin u)^\nu-u^\nu]+\frac{1}{\nu+1}\left(
\frac{\pi}{2}\right)^{\nu+1}=\frac{\sqrt{\pi}}{2}\frac{\Gamma\left(
\frac{1+\nu}{2}\right)}{\Gamma(1+\nu/2)}.
\end{equation}

\item
From the recursion (\ref{recf}) and (\ref{fam1}), we can calculate the value of
the coefficients  
\begin{equation}
f_0=2^{-d}\frac{\Gamma(\Delta-1)(-1)^a\sqrt{\pi}}{\Gamma\left(
\frac{d-1}{2}\right)\Gamma(\nu+2)\pi^{d/2}},\qquad  
\pi c=2^{-d}\frac{\Gamma(\Delta)(-1)^a\sqrt{\pi}}{2\pi^{d/2}\Gamma(D/2)
\Gamma(\nu+1)}.  
\end{equation}

\item
Using the above result, we have 
\begin{equation}
\pi c G_o(2\Delta-d)=2^\nu\xi.
\end{equation}
\end{itemize}




\bibliographystyle{JHEP}
\bibliography{Flow}

\end{document}